\documentclass[fleqn,usenatbib]{mnras}

\usepackage{newtxtext,newtxmath}
\usepackage{fontawesome}

\usepackage[T1]{fontenc}

\DeclareRobustCommand{\VAN}[3]{#2}
\let\VANthebibliography\thebibliography
\def\thebibliography{\DeclareRobustCommand{\VAN}[3]{##3}\VANthebibliography}

\usepackage{graphicx}	
\usepackage{amsmath}	

\title[Method to constrain the EoR's History]{Using Neural Emulators and Hamiltonian Monte Carlo to constrain the Epoch of Reionization's History with the Ly$\alpha$ Forest Power Spectrum}

\author[D. González-Hernández et al.]{
Diego González-Hernández,$^{1}$\thanks{E-mail: dgonzalezhernandez@ucsb.edu}
Caitlin Doughty,$^{2}$
Molly Wolfson,$^{3,4,5}$
Joseph F. Hennawi, $^{1,6}$
Zhenyu Jin $^{7,8}$
\\
$^{1}$Department of Physics, University of California, Santa Barbara, CA 93106, USA\\
$^{2}$Max-Planck-Institut für Astrophysik, Karl-Schwarzschild-Str. 1, D-85748 Garching, Germany\\
$^{3}$Center for Cosmology and AstroParticle Physics, The Ohio State University, 191 West Woodruff Avenue, Columbus, OH 43210, USA\\
$^{4}$Department of Physics, The Ohio State University, 191 West Woodruff Avenue, Columbus, OH 43210, USA\\
$^{5}$Department of Astronomy, The Ohio State University, 4055 McPherson Laboratory, 140 W 18th Avenue, Columbus, OH 43210, USA\\
$^{6}$Leiden Observatory, Leiden University, Niels Bohrweg 2, 2333 CA Leiden, Netherlands\\
$^{7}$Department of Physics, University of Washington, 1410 NE Campus Pkwy, Seattle, WA 98195, USA\\
$^{8}$Berkeley Center for Cosmological Physics, Department of Physics, University of California, Berkeley, 341 Campbell Hall, Berkeley, CA 94720, USA
}

\date{Accepted XXX. Received YYY; in original form ZZZ}

\pubyear{\the\year{}}

\begin{document}
\label{firstpage}
\pagerange{\pageref{firstpage}--\pageref{lastpage}}
\maketitle

\begin{abstract}
The Lyman-alpha (Ly$\alpha$) forest at $z \sim 5$ offers a primary probe to constrain the history of the Epoch of Reionization (EoR), retaining thermal and ionization signatures imprinted by the reionization process. In this work, we present a new inference framework based on JAX that combines forward-modeled Ly$\alpha$ forest observables with differentiable neural emulators and Hamiltonian Monte Carlo (HMC). We construct a dataset of 501 low-resolution simulations generated with user-defined reionization histories and compute a set of 1D Ly$\alpha$ power spectra and model-dependent covariance matrices. We then train two independent neural emulators that achieve sub-percent errors across relevant scales and combine them with HMC to efficiently perform parameter estimation. We validate this framework by applying it to a suite of mock observations, demonstrating that the true parameters are reliably recovered. While this work is limited by the low resolution of the simulations used, our results highlight the potential of this method for inferring the reionization history from high-redshift Ly$\alpha$ forest measurements. Future improvements in our reionization models will further enhance its ability to extract constraints from  observational datasets.
\end{abstract}




\section{Introduction}
The Epoch of Reionization (EoR) marks a major transition in the universe's history during which the first luminous sources ionized the neutral hydrogen in the intergalactic medium (IGM) \citep{Loeb2001, Gnedin2022}. 
Currently, the best constraints on the EoR's history come from the integrated Thomson scattering optical depth $\tau_{\text{e, CMB}}$ obtained from measurements of the Cosmic Microwave Background (CMB). These constraints paint a scenario where the EoR took place during redshift range between $z = 6.4 -9.0$, with a midpoint at around $z_{\text{mid}}=7.7\pm0.7$ \citep{PlanckCollaboration2020b, PlanckCollaboration2020}. However, because $\tau_{\text{e, CMB}}$ is an integral constraint, it is insensitive to the detailed duration and inhomogeneous structure of the EoR \citep{Doughty2025}, leaving its precise history poorly constrained.
Future missions and facilities will seek to improve these constraints by using the 21 cm line \citep[e.g.][]{DeBoer2017}, creating tomographic maps of the neutral hydrogen throughout the EoR. However, there is still a significant amount of work left for these techniques to properly work, with current efforts focusing on obtaining a clean signal from multiple sources of contamination.

The next best constraints come from a set of high-$z$ quasar and galaxy measurements. Specifically, observations of damping wings in quasar spectra have been used to measure the neutral hydrogen fraction at redshifts $7 < z < 7.5$ \citep[e.g.][]{Greig2017, Banados2018, Davies2018, Greig2019, Wang2020, Yang2020, Greig2022}. These measurements, however, yield a broad range of inferred neutral fractions, likely reflecting differences in the underlying inference techniques, though recent methodological developments offer promising avenues to mitigate these discrepancies \citep{Durovcikova2024, Hennawi2025, Kist2025b, Kist2025}. Similar approaches applied to galaxies have provided additional constraints \citep[e.g.][]{Mason2019}, but they are limited by uncertainties in continuum modeling \citep{Keating2024} and potential contamination from neutral hydrogen in galaxies’ circumgalactic and intergalactic media \citep{Heintz2024}, reducing their robustness.

The Lyman-$\alpha$ (Ly$\alpha$) forest is a primary probe of neutral hydrogen during the final stages of the EoR, tracing absorption along quasar sightlines \citep{Gunn1965, Lynds1971}. Building on this framework, \citet{McGreer2015} used the ``dark pixel'' fraction in quasar spectra to place a robust, model-independent upper limit of $x_{\mathrm{HI}} < 0.09$ at $z=5.6$. Additional constraints have been derived from Ly$\alpha$ and Ly$\beta$ gaps, though these results are more model-dependent \citep{Zhu2022, Jin2023}. Achieving tighter limits would require leveraging the full distribution of Ly$\alpha$ opacities, making improved statistical treatments of the Ly$\alpha$ forest increasingly important for constraining residual neutral hydrogen at $z<6$.

A key example of this is the Ly$\alpha$ flux power spectrum analysis of \citet{Boera2019}, where high-resolution quasar spectra at $4 \lesssim z \lesssim 5.2$ were used to measure IGM temperatures of $T_0 \sim 7000$–$8000$ K and place the first constraints on its integrated thermal history. Their results indicated ongoing photoheating after reionization and favored a relatively late reionization scenario, with an inferred midpoint of $z_{\mathrm{rei}} \approx 8.5^{+1.1}_{-0.8}$, broadly consistent with CMB optical depth constraints. Additionally, different analyses using clustering measures of the Ly$\alpha$ forest have reinforced the picture of reionization completing near $z \sim 6$ while highlighting the potential of the Ly$\alpha$ forest to probe both residual neutral hydrogen and the thermal imprint of reionization. \citep{Fan2006, Becker2015, Bosman2018, Eilers2018, Bosman2022}. 

While these constraints are promising, a more detailed description of the reionization histories is not present in the models used in the cited studies, which prevents a tighter constraint on the EoR. To tackle this problem, \citet{Doughty2025} proposed the use of models in which the reionization history is the primary focus. To achieve this, a combination of AMBER \citep[a code used to generate reionization fields that correspond to user-defined reionization histories, see Section \ref{subsec:AMBER} and][]{Trac2022} and hydrodynamical simulations was used to generate a variety of metrics of the Ly$\alpha$ forest, demonstrating that these are sensitive to the EoR's history. 
However, a typical simulation with enough resolution to correctly capture the relevant small scale structures can require thousands of GPU-hours \citep{Doughty2023a}. Because of this, \citet{Doughty2025} was limited to a qualitative comparison with observational data. 

To overcome this challenge, a viable option is to construct appropriate emulators that act as computationally cheap surrogates of the full models, a technique widely used in cosmology \citep{Heitmann2009, Kwan2015, Liu2015, Petri2015, Jennings2019, McClintock2019, Zhai2019, Hennawi2025}. A variety of techniques have been developed to emulate the Ly$\alpha$ forest. Earlier efforts focused on methods such as Taylor expansions and quadratic polynomial interpolation \citep{Viel2006, Bird2011, Palanque2013, Palanque2015, Yeche2017, Palanque2020}, 
as well as Gaussian process models 
\citep{Rorai2013, Bird2019, Rogers2019, Walther2019, Walther2021, Pedersen2021, Rogers2021, Fernandez2022, Bird2023}.

More recently, the rise of deep learning has enabled the use of neural networks as Ly$\alpha$ forest emulators. Several applications have explored this direction. For instance, \citet{Huang2021} trained a neural network to estimate the neutral hydrogen density from Ly$\alpha$ transmission flux, while \citet{Harrington2022} used a convolutional neural network to generate baryonic hydrodynamic variables on Ly$\alpha$-relevant scales from N-body simulations. Other examples include reconstructing the IGM temperature from Ly$\alpha$ flux \citep{Wang2022}, emulating the 1D power spectrum for cosmological parameter inference at $z\sim2$–5 \citep{Cabayol-Garcia2023, Molaro2023}, and predicting IGM gas properties such as temperature and optical depth from Ly$\alpha$ skewers \citep{Nasir2024}. Further recent work has applied deep learning for field-level inference of thermal parameters from Ly$\alpha$ observations at $z=2.2$ using residual networks \citep{Nayak2024}, estimating thermal parameters in Fourier space via information-maximizing networks \citep{Maitra2024}, and emulating the Ly$\alpha$ forest autocorrelation function to accelerate inference at $z\sim5.4$–6.0 \citep{Zhenyu2025}.

An advantage of constructing neural emulators is the possibility of taking derivatives of their output with respect to their input, calculated via automatic differentiation. This capability plays a central role in the backpropagation algorithm used to train neural networks \citep{Rumelhart1986, Baydin2015}. Taking advantage of this, some studies have explored the possibility of combining neural emulators with gradient-based sampling algorithms to accelerate parameter estimation. In particular, \citet{Piras2023} created differentiable emulators of cosmological power spectra which were combined with Hamiltonian Monte Carlo (HMC) to accelerate cosmological inference. Following a similar approach, \citet{Zhenyu2025} combined a neural emulator with HMC to accelerate the inference of the IGM's thermal parameters, while \citet{Carrion2025} followed the same approach to forecast constraints on dark energy models from cosmic shear surveys. Beyond this specific approach, other studies have also explored the possibility of creating fully differentiable modeling software to accelerate parameter estimation with gradient-based sampling algorithms. These include a library for fully automatic differentiable cosmological theory calculations \citep{Campagne2023}, a forward modeling framework for line intensity mapping cosmology experiments focused on 21 cm cosmology \citep{Kern2025}, using automatic differentiable programming to solve the equation of state in neutron stars \citep{Wouters2025}, and a fully differentiable hydrodynamics code to perform field-level inference of cosmological initial conditions \citep{Horowitz2025}.

Another important requirement for parameter estimation is the proper calculation of covariance matrices. Typically, covariance matrices can be estimated directly from observational data when large samples are available. Unfortunately, towards the end of the EoR ($z=5$–$6$), the number of available quasar sightlines within narrow redshift bins is severely limited ($\sim$10–20), making it difficult to robustly estimate the true covariance matrix. This limitation arises because high-$z$ quasars are intrinsically rare and faint. As a result, even the largest high-resolution samples to date \citep[such as XQR-30, which includes 30 quasars at $5.8 \lesssim z \lesssim 6.6$,][]{D'Odorico2023}, remain too small to empirically derive well-converged covariance matrices.

To solve this, \citet{Boera2019} constructed a covariance matrix by combining simulations and data: they computed a correlation matrix from a single hydrodynamical simulation (assuming fixed cosmological and thermal parameters), and then rescaled its diagonal using an empirical variance estimated from bootstrap resampling of their 15 observed quasar spectra. This approach provided a stable covariance estimate but implicitly assumed that the correlation structure from one fiducial simulation applies across parameter space. Additionally, this covariance was held fixed during inference, potentially ignoring the effects that different thermal histories could have on the structure of the covariance matrix. Alternatively, fully model-dependent covariance matrices can be computed directly from large ensembles of carefully forward-modeled mock observations \citep[e.g.][]{Wolfson2023, Wolfson2023mfp}. While this approach captures parameter dependence explicitly, it is computationally prohibitive, typically requiring an entirely new simulation to generate a covariance matrix. Here, once again, emulators could in principle alleviate the high computational costs. Unfortunately, emulating matrices is far more challenging than emulating vector-valued statistics, as it requires correctly approximating a smooth mapping between the model parameters and the structured, positive semi-definite matrices across all the parameter space of interest. Importantly, no previous studies have attempted neural network emulation of Ly$\alpha$ forest covariance matrices,

In this work, we address these challenges and develop a method that will allow a quantitative constraint of the EoR's history using the 1D power spectrum of the Ly$\alpha$ forest at a redshift of $z=5.0$. We start by creating a suite of $501$ simulations that are used to generate a dataset of model-dependent power spectra and covariance matrices in Section \ref{sec:models}. We continue by training independent neural emulators for the power spectra and covariance matrices, and combine them with HMC to perform accelerated parameter inference in Section \ref{sec:parameter-inference}. In Section \ref{sec:results} we evaluate the performance of our emulators and the statistical validity of the posterior distributions we obtain from performing parameter inference on a set of mock observations. We finalize by discussing our results and providing our conclusions in Section \ref{sec:conclusions}. Unless otherwise stated, all distances in this work are quoted in comoving units.

\section{Reionization models}\label{sec:models}

The evolution of the $\text{Ly}\alpha$ forest during the Epoch of Reionization can be modeled using different techniques that vary in physical fidelity and computational cost. Fully coupled radiation-hydrodynamical simulations offer the most accurate and physically motivated approach, as they solve the radiative transfer equations in situ while allowing the gas to thermodynamically respond to ionizing radiation \citep[e.g.][]{Gnedin2014, Ocvirk2016, Finlator2018, Rosdahl2018, Rosdahl2022, Kannan2022}. However, these simulations are computationally expensive 
and typically limited to small cosmological volumes, restricting their use for statistical studies of large-scale reionization signatures \citep{Trac2022, Doughty2023a}. 
An intermediate approach relies on partially coupled or hybrid simulations, where the radiation field and gas dynamics are evolved separately, typically by applying radiative transfer in post-processing or by using simplified source models \citep[e.g.][]{Iliev2006, Bauer2015, Onorbe2017, Eide2018, Huang2018, Onorbe2019, Bird2022, Puchwein2023, Doughty2025}.

Alternatively, semi-numerical methods offer an efficient way to generate large-scale reionization fields by applying analytical approximations to evolved density distributions, often leveraging the excursion set formalism to model the expansion of ionized bubbles \citep[e.g.][]{Mesinger2011, Battaglia2013, Davies2016, Feng2016, Trac2022}. 
These models can operate independently or in conjunction with hydrodynamical simulations, making them especially useful for exploring broad parameter spaces with relatively low computational requirements. In some cases, dark matter-only simulations are combined with semi-analytical prescriptions for ionizing sources, providing a flexible platform for modeling ionization fields across different reionization histories \citep{Ghara2015, Poole2016, Hutter2021}. Given our goal of constraining the global reionization history, we base our models on AMBER \citep{Trac2022}, a semi-numerical framework capable of producing thermal and ionization fields corresponding to user-defined reionization histories. We describe this method in detail in the following subsection.

\subsection{AMBER reionization histories}\label{subsec:AMBER}

\begin{figure}
    \includegraphics[width=\columnwidth]{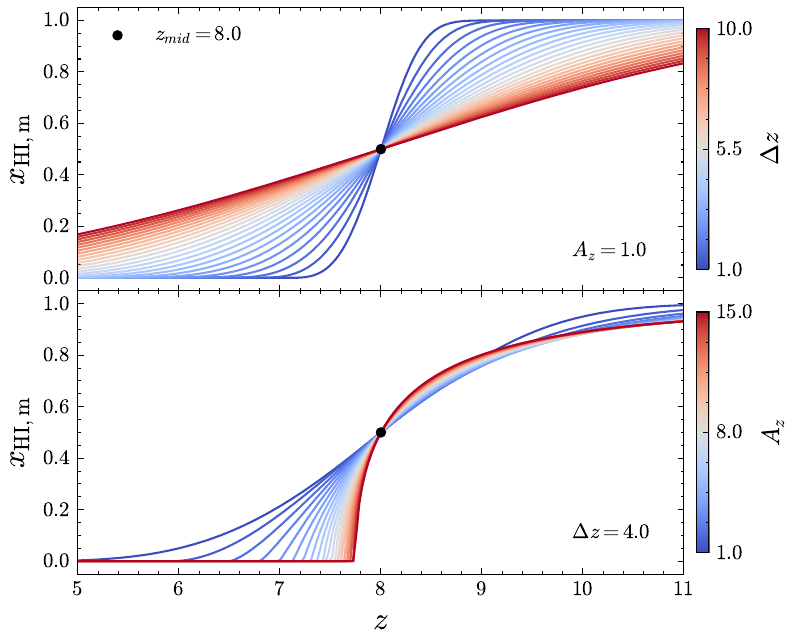}
    \caption{Examples of different mass-weighted reionization histories generated with AMBER's analytical model. \textit{Top Panel:} demonstrates the effect of varying the duration $\Delta z$, while fixing $z_{\text{mid}}=8.0$ and $A_z=1.0$. \textit{Bottom Panel:} shows the influence of the asymmetry parameter $A_z$, with fixed $z_{\text{mid}}=8.0$ and $\Delta z=4.0$.}
    \label{fig:reion-histories-examples}
\end{figure}

The Abundance Matching Box for the Epoch of Reionization (AMBER) is a semi-numerical code designed to model the Cosmic Dawn and the Epoch of Reionization \citep{Trac2022}. It constructs a gridded reionization field using an abundance-matching scheme that assumes hydrogen gas exposed to stronger ionizing radiation is reionized earlier. One of AMBER's key innovations is its ability to generate ionization fields corresponding to a user-defined reionization history. This is achieved by describing the evolution of the 
\textit{mass-weighted}\footnote{While AMBER is built around the \textit{mass-weighted} ionized fraction (appropriate for tracking ionized mass density), it also naturally produces a corresponding (and more typically used) \textit{volume-weighted} ionized fraction, $x_{\mathrm{HII,V}}(z)$, which is lower at fixed redshift since higher-density regions tend to ionize first \citep[for more details, see][]{Trac2022}.} ionized hydrogen fraction, $x_{\text{HII,m}}(z)$, using three parameters: the midpoint of reionization $z_{\text{mid}}$, its duration $\Delta z$, and its asymmetry $A_z$. The midpoint is defined such that $x_{\text{HII,m}}(z_{\text{mid}}) = 0.5$. The duration is given by:
\begin{equation}
\Delta z \equiv z_{\text{ear}} - z_{\text{late}}    
\end{equation}
where $z_{\text{ear}}$ and $z_{\text{late}}$ are defined such that $x_{\text{HII,m}}(z_{\text{ear}}) = 0.05$ and $x_{\text{HII,m}}(z_{\text{late}}) = 0.95$. Lastly, the asymmetry is defined as:
\begin{equation}\label{eq:asymmetry}
A_z \equiv \frac{z_{\text{ear}} - z_{\text{mid}}}{z_{\text{mid}} - z_{\text{late}}}
\end{equation}
By this definition, $A_z$ quantifies the relative duration of the early and late phases of reionization. Intuitively, using values $A_z > 1$ correspond to histories with a slower initial rise in ionization followed by a rapid completion. Once the parameters $z_{\text{mid}}$, $\Delta z$, and $A_z$ are specified, AMBER computes the corresponding early and late reionization redshifts using:
\begin{equation}
z_{\text{ear}} = z_{\text{mid}} + \frac{\Delta z A_z}{1 + A_z}
\end{equation}
\begin{equation}
z_{\text{late}} = z_{\text{mid}} - \frac{\Delta z A_z}{1 + A_z} = z_{\text{ear}} - \Delta z
\end{equation}
AMBER then interpolates between these three redshift points using a modified Weibull function \citep{Weibull1951}, yielding a smooth and flexible evolution for $x_{\text{HII,m}}(z)$. This parametrization has been shown to closely reproduce reionization histories extracted from full radiation-hydrodynamic simulations \citep{Trac2018}. Figure~\ref{fig:reion-histories-examples} shows examples of the mass weighted reionization histories that can be generated with AMBER, and the effects of varying both $\Delta z$ and $A_z$.

In this work, we apply AMBER in a similar way as described in \citet{Doughty2025}. In summary, the density field is generated from a Gaussian random field, that is first calibrated to the cosmological linear matter power spectrum. AMBER allows for either the Zel’dovich approximation \citep{Zel'dovich1970}, or the second-order LPT \citep[2LPT][]{Bouchet1995, Scoccimarro1998}. In our case, we use 2LPT to evolve the initial conditions to $z = z_{\mathrm{mid}}$. The evolved particle distribution is then mapped to a grid which is used for the simulations. While AMBER can internally generate its own Gaussian random field, we bypass this functionality and instead extract the $z = 0$ density field from an independent initial conditions generator to ensure consistency with our hydrodynamical simulations \citep[see][for more details]{Doughty2025}.

After selecting the reionization parameters and generating the density field, AMBER computes the ionizing radiation field using an excursion set-based approach. This involves estimating the local halo density from the overdensity distribution, applying the extended Press–Schechter formalism \citep{Bond1991, Lacey1993}. The ionizing background is then modeled using a simplified scheme that combines a prescribed source function with a fixed photon mean free path, which we set to 3 $h^{-1}$ Mpc.
following \citet{Doughty2025}. The source function incorporates astrophysical parameters like the escape fraction and star formation efficiency, which are adjusted to produce the target reionization history \citep[e.g.][]{Iliev2006}. After computing the spatial distribution of radiation, AMBER rank-orders the simulation cells by radiation intensity and maps them to the cumulative ionized mass fraction, $x_{\mathrm{HII},\mathrm{m}}(z)$. This procedure assigns a reionization redshift to each cell, producing a reionization field that captures both the large-scale topology and the specified global history. This approach enables the efficient generation of self-consistent reionization fields, making AMBER well suited for parameter studies such as the one presented in this work.

\subsection{Nyx simulations}\label{subsec:nyx-sims}
Our simulations are run with Nyx \citep{Almgren2013, Sexton2021}, a widely used hydrodynamics code for cosmological modeling that has been applied extensively in studies of the $\text{Ly}\alpha$ forest \citep[e.g.][]{Lukic2015, Onorbe2019, Chabanier2023, Wolfson2023, Jacobus2023}. As in \citet{Doughty2025}, we compute the necessary transfer functions with CAMB \citep{Lewis2000} and generate the initial conditions using CICASS \citep{OLeary2012}, including a baryon–dark matter streaming velocity of $\langle \Delta v \rangle = 30$ km/s at recombination. Initial conditions are generated at $z = 200$ and used to construct the overdensity field input for AMBER, following the approach of \citet{Doughty2025}. 

In this work, we use a simulation box size of $L_{\text{box}} = 20$ $h^{-1}$ Mpc, and a resolution of $256^3$ voxels. This is in the low-resolution regime, and is not able to fully resolve the small scales that capture important effects of the EoR on the Ly$\alpha$ forest \citep{Doughty2023a}. However, since the main objective of this study is to develop our emulation and inference methods, 
this resolution is sufficient for our goal. For the simulations, we assume a $\Lambda$CDM cosmology consistent with results from Planck \citep{PlanckCollaboration2020}, ($\Omega_M$, $\Omega_\Lambda$, $\Omega_b$, $h$, $X_H$ ) = ($0.315$, $0.685$, $0.049$, $0.675$, $0.76$).

Since Nyx does not model galaxy formation, and feedback effects have been shown to have minimal impact on the $\text{Ly}\alpha$ forest at high redshifts \citep{Viel2013b}, we choose to exclude any feedback prescriptions from our simulations. Thus, the thermal state of the gas is mainly driven by the reionization event (assumed to the photonoionization of a given cell by a uniform UV background after it reaches its reionization redshift), 
structure formation, and the associated shock heating. 

In Nyx, once a cell reaches its reionization redshift (as determined from AMBER’s reionization-redshift field), it is heated to account for the passage of an ionization front. This is implemented by injecting an amount of heat $\Delta T$ that depends on the cell’s pre-reionization gas temperature $T_{\mathrm{pre}}$ and neutral hydrogen fraction $x_{\mathrm{H,I,,pre}}$, following \citet{Doughty2025}:
\begin{equation}
T_{\mathrm{post\text{-}reion}} = \left(x_{\mathrm{H\,I,\,pre}}\right) \max\left(\Delta T_{\mathrm{re}} - T_{\mathrm{pre}},\, 0 \right) + T_{\mathrm{pre}}.
\end{equation}
Here, $\Delta T_{\mathrm{re}}$ is the user-defined maximum amount of heating that can be injected (typically $10{,}000$–$20{,}000$ K). This heating accounts for the energy deposited by photoionization and the associated spectral hardening of the radiation field as the ionization front propagates \citep[see][]{D'Aloisio2019}. 

\begin{figure*}
    \centering
	\includegraphics[width=1.95\columnwidth]{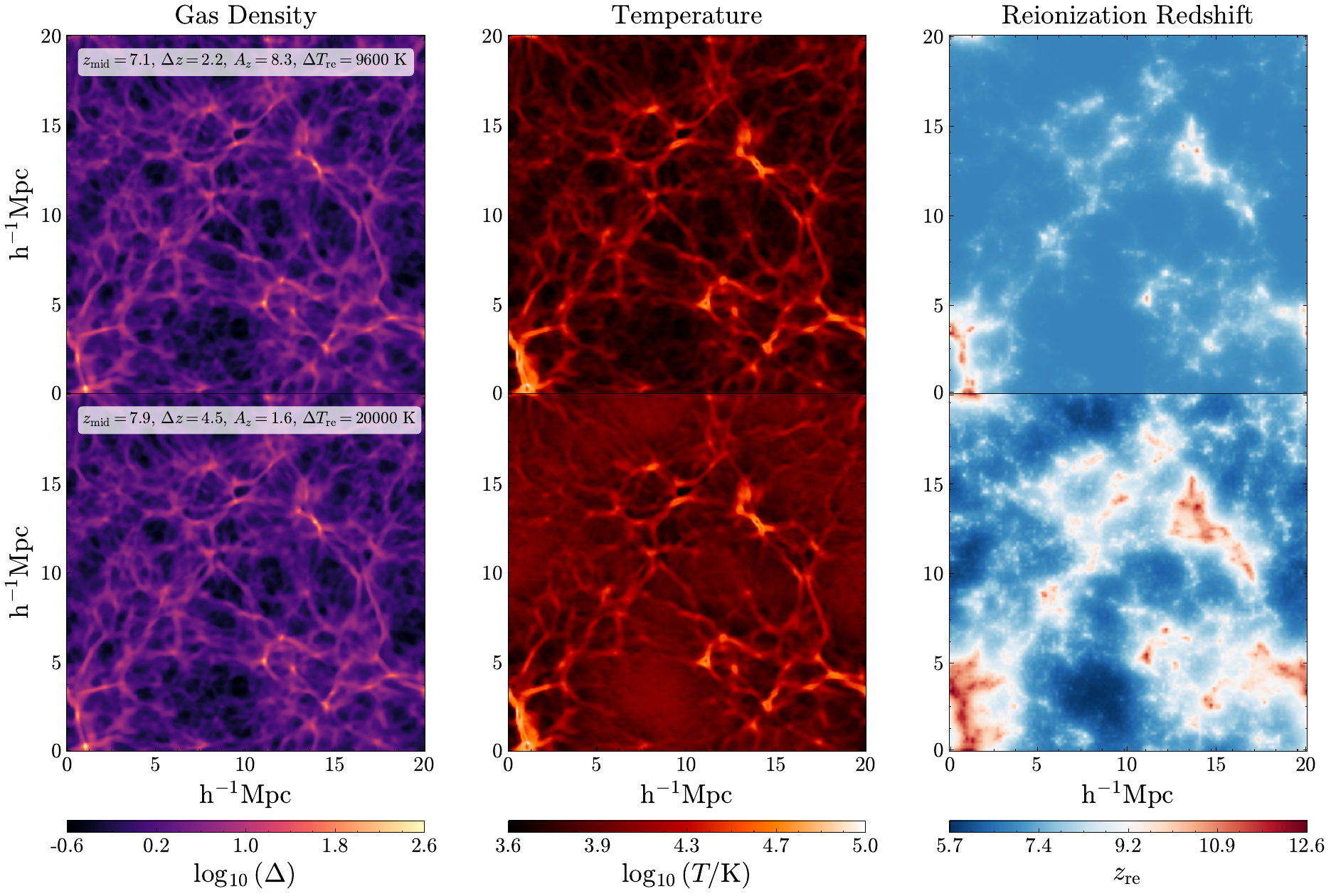} 
    \caption{Slices of 88.5 km/s at $z=5.00$ of 20 $h^-1$ cMpc from the simulations. The left column shows the gas overdensity field ($\Delta$), while the center and right columns show the temperature ($T$) and reionization redshift ($z_{\text{re}}$) fields, respectively. Since the simulations have initial conditions with the same seed, the overdensity structures remain the same, and the differences in the temperature and reionization redshift fields come from the simulation's reionization histories. The slices in the top row correspond to Model 1, corresponding to a reionization history with parameters $z_{\mathrm{mid}} = 7.1$, $\Delta z = 2.2$, $A_z = 8.3$, $\Delta T_{\mathrm{re}} = 9600$ K. The bottom row corresponds to Model 2, with parameters $z_{\mathrm{mid}} = 7.9$, $\Delta z = 4.5$, $A_z = 1.6$, $\Delta T_{\mathrm{re}} = 20000$ K. Their corresponding volume weighted reionization histories are shown in Figure \ref{fig:filtered_reion_histories}.}
    \label{fig:simulation-slice}
\end{figure*}

Figure \ref{fig:simulation-slice} shows 88.5 km/s slices of the gas density and temperature fields, as well as their corresponding reionization redshift field, at a redshift of $z=5.00$ taken from two distinct 20 $h^{-1}$ Mpc simulations. Both simulations have the same initial conditions and present the same structure formation and gas overdensities as a consequence. However, the temperature and reionization fields present clear differences, as these depend on the reionization histories. Model 1 (shown in the top row) corresponds to a simulation with parameters $z_{\mathrm{mid}} = 7.1$, $\Delta z = 2.2$, $A_z = 8.3$, $\Delta T_{\mathrm{re}} = 9600$ K, which correspond to a late, rapid, and asymmetric reionization history 
with cool ionization fronts. In contrast, Model 2 (shown in the second row) corresponds to a simulation with parameters  $z_{\mathrm{mid}} = 7.9$, $\Delta z = 4.5$, $A_z = 1.6$, $\Delta T_{\mathrm{re}} = 20000$ K, corresponding to an earlier, longer, and symmetric reionization history with hot ionization fronts.

\subsection{Forward Modeling}\label{subsec:forward-modeling}
For a given simulation snapshot, we use Nyx's output to generate $N_{\text{skwrs}}$ optical depth skewers ($\tau$-skewers). Given that our models do not directly model the average of the UVB, we have introduced an extra parameter, the mean flux $\langle F \rangle$ as a proxy for the intensity of the UVB. This is done by computing flux skewers ensuring that $\langle e^{-\tau} \rangle = \langle F \rangle$ when averaging the transmitted flux over all the skewers for a user-defined value of $\langle F \rangle$. This leaves us with $N_{\text{skwrs}}$ perfect (i.e. noiseless) flux skewers.

In order to mimic realistic observational data from echelle spectrographs (such as Keck/HIRES, VLT/UVES, and Magellan/MIKE), we forward model our perfect flux skewers to include resolution effects and noisy flux levels. We consider a resolution of $R = 30000$ and a signal to noise ratio per 10 km s$^{-1}$ region of
$\text{SNR}_{10} = 30$. To account for the resolution, we smooth the flux using a Gaussian filter with FWHM = 10 km s$^{-1}$  and then re-sample the new flux onto a uniform velocity grid with pixel size  $\Delta v = 2.5$ km s$^{-1}$. 

\begin{figure}
    \includegraphics[width=\columnwidth]{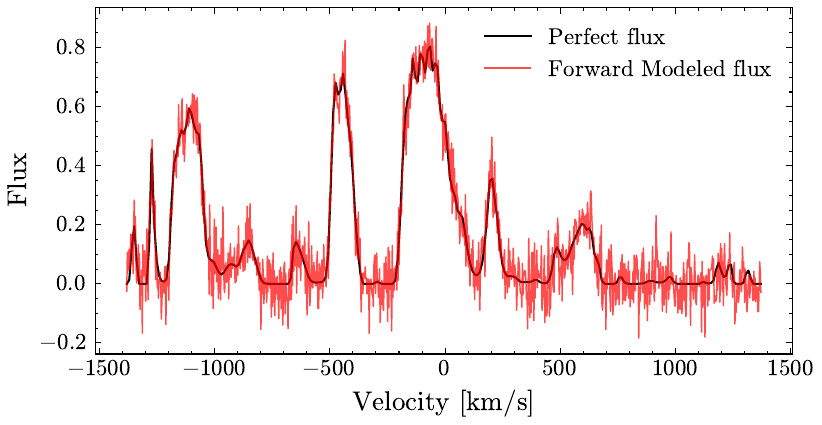}
    \caption{An example of a perfect skewer and its corresponding forward modeled skewer at $z=5.0$, taken from a model with parameters $z_{mid}$ = 8.9, $\Delta z$ = 2.0, $A_{z}$ = 2.4, $\Delta T_{\text{re}}$ = 4100 K, $\langle F \rangle$ = 0.1845. All forward modeling is done assuming $R=30000$ and $\text{SNR}_{10} = 30$.}
    \label{fig:forward-model}
\end{figure}

At this pixel scale, $\text{SNR}_{10} = 30$ corresponds to a signal to noise ratio per $\Delta v = 2.5$ km s$^{-1}$ pixel of $\text{SNR}_{\Delta v} = 15$. We assume the noise is Gaussian and add random noise to each model's flux skewers in the following way: we generate one $N_{\text{skwrs}}$ $\times$  skewer length realization of random noise drawn from a Gaussian with $\sigma_N = 1/\text{SNR}_{\Delta v}$ and use this exact noise realization to every model we consider. Using the same noise realization over the different models prevents stochasticity from different realizations of the noise from adding additional variations between the models, while still allowing us to properly forward model the inclusion of noise. Figure \ref{fig:forward-model} shows an example of a typical forward modeled skewer.

\subsection{Model dependent products}\label{subsec:model-products}
To be able to test our method, we compute a 1D power spectrum of the Ly$\alpha$ forest, mock observations, and a covariance matrix for each model. We now describe how each one of these products are computed.

\subsubsection{Mean Power Spectrum}\label{subsec:mean-power-spectrum-model}
\begin{figure}
    \includegraphics[width=\columnwidth]{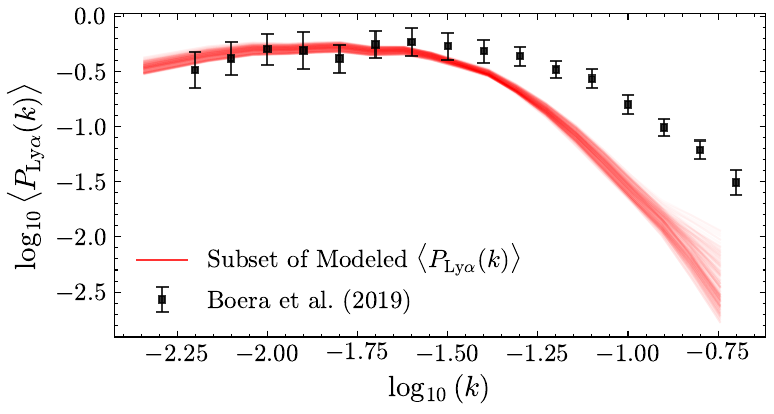}
    \caption{An subset of 200 randomly selected $\langle P_{\mathrm{Ly\alpha}} \rangle$ from our model compared to the measurement from \citet{Boera2019} at $z=5.0$. These models match the $\langle F \rangle$ from the observational data. As shown, the suppression of power at smaller scales is present in all the models we compute, and is caused by the low resolution of our simulations (see Section \ref{subsec:mean-power-spectrum-model}).}
    \label{fig:power-spectrum-boera}
\end{figure}

The dimensionless mean 1D power spectrum of the Ly$\alpha$ forest, $\langle P_{\mathrm{Ly\alpha}} \rangle$, is a summary statistic that has been used extensively in the literature to constrain cosmology and characterize the IGM \citep[e.g.][]{McDonald2006, Palanque2013, Palanque2015, Chabanier2019, Walther2019, Boera2019}. 
For a given flux skewer, the power spectrum $P_{\mathrm{Ly\alpha}}$ is defined by:
\begin{equation}
P_{\mathrm{Ly\alpha}}(k) = \delta_F(k)^*\delta_F(k)
\end{equation}
Where $\delta_F(k)$ is the Fourier transform of the flux overdensity, defined as:
\begin{equation}
	\delta_F \equiv \frac{F-\bar{F}}{\bar{F}}
\end{equation}
Where $F$ is the flux in units of the continuum level, and $\bar{F}$ is the mean flux of the individual flux skewer. For a given model, we estimate $\langle P_{\mathrm{Ly\alpha}} \rangle$ by first calculating the individual power spectra from each perfect flux skewer, and then averaging them. We use logarithmically spaced $k$-bins, where each $k$ value indicates the center of the bin. We use 19 evenly separated $\log_{10}(k)$ values that range from $-2.34 < \log_{10}(k)<-0.74$.  Figure \ref{fig:power-spectrum-boera} shows a set of 200 randomly selected power spectra $\langle P_{\mathrm{Ly\alpha}} \rangle$ generated with our model compared to the measurement from 
\citet{Boera2019} at $z=5.0$. For these examples, we use the same value of $\langle F \rangle$ = 0.1845 used in \citet{Boera2019},
and use random combinations of values for the rest of the parameters, each within their respective ranges shown in Table \ref{tab:parameter-ranges}. As we can see, our modeled $\langle P_{\mathrm{Ly\alpha}} \rangle$ is not able to match the power observed at smaller scales. This is because our simulations are low resolution, and they do not fully resolve the small scale structure, causing a suppression of power at these scales. Since the main objective of this study is to develop and test the approach of using fully differentiable emulators for both the mean power spectrum and the covariance matrices, 
we decide to proceed with these low resolution simulations, and leave the creation of a high resolution dataset of models and its application to observational measurements for future work.

\subsubsection{Mock Observations}
For each model, we generate a set of a million $N_{\text{mocks}}$ mock observations of the $P_{\mathrm{Ly\alpha}}$. Following the standard approach \citep[as seen in e.g.][]{Palanque2013, Boera2019}, we calculate an individual mock observation $P_{\mathrm{Ly\alpha}, i}$ using the following:
\begin{equation}
P_{\mathrm{Ly\alpha}, i} = \frac{|P_{\text{raw}} - P_N|}{W^2(k, R, \Delta v)}
\end{equation}
Where $P_{\text{raw}}$ is the mean power spectrum calculated from a subset of randomly selected forwarded modeled skewers, $P_N$ is the noise power spectrum, and $W^2$ is a Gaussian window function (which corresponds to the resolution of a spectrograph) adopted in
\citet{Palanque2015}. We assume that each mock observation corresponds to a measurement performed on a dataset of 20 quasar sightlines that probe a redshift interval $dz = 0.1$. Centering this interval at $z=5.0$ (see Section \ref{subsec:simulation-dataset}), 20 quasar sightlines corresponds to $\sim53$ flux skewers from our simulations, given our value of $L_{\text{box}}$ (see Section \ref{subsec:nyx-sims}). Thus, we use 53 forward modeled flux skewers to calculate each mock observation $P_{\mathrm{Ly\alpha}, i}$. These mock observations are necessary to compute the covariance matrices (see Section \ref{subsec:covar-matrices}) and to test our inference approach (see Section \ref{sec:parameter-inference}).

\subsubsection{Covariance Matrices}\label{subsec:covar-matrices}
\begin{figure}
    \includegraphics[width=0.87\columnwidth]{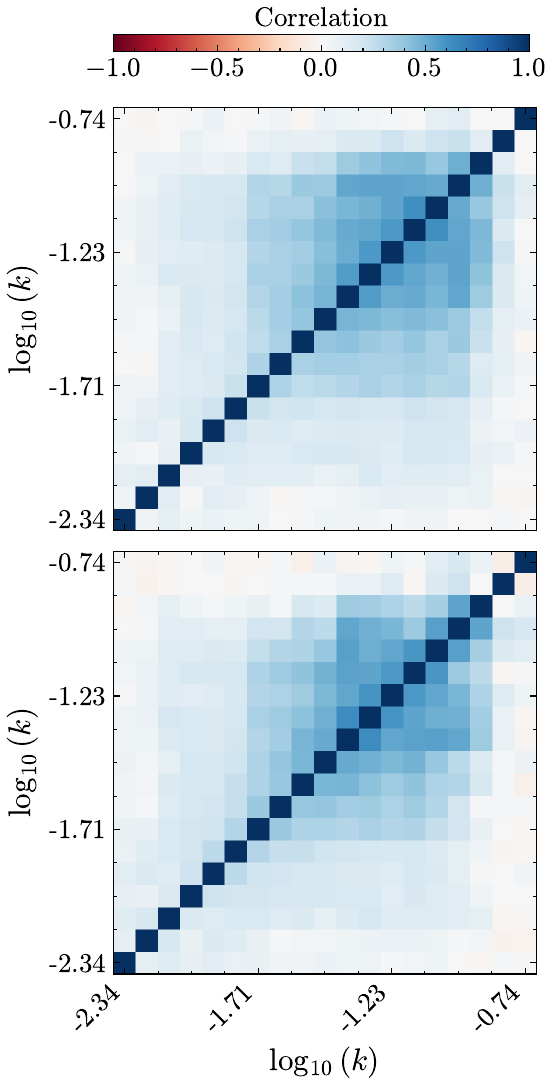}
    \caption{Examples of a covariance matrices (shown as a correlation matrices) at $z=5.0$. The covariance matrix at the top corresponds to a model with $z_{\text{mid}}$ = 6.4, $\Delta z$ = 1.0, $A_{z}$ = 12.4, $\Delta T_{\text{re}}$ = 29000, $\langle F \rangle$ = 0.1496. The covariance matrix at the bottom corresponds to a model with $z_{\text{mid}}$ = 8.7, $\Delta z$ = 16.0, $A_{z}$ = 11.5, $\Delta T_{\text{re}}$ = 36000, $\langle F \rangle$ = 0.1156.}
    \label{fig:covar-examples}
\end{figure}

Given our assumption of a multivariate Gaussian likelihood function (see Section \ref{sec:parameter-inference}), we generate covariance matrices $\Sigma_{P_{\mathrm{Ly\alpha}}}$. For each model, we use the corresponding mock observations to compute $\Sigma_{P_{\mathrm{Ly\alpha}}}$ in the following manner\footnote{Although the unbiased estimator for a sample covariance matrix typically uses $N_{\text{mocks}} - 1$ in the denominator, we use $N_{\text{mocks}}$ here because the mean $\langle P_{\mathrm{Ly\alpha}} \rangle$ is computed separately from noiseless skewers rather than from the same mock realizations used to estimate the covariance.}:
\begin{equation}\label{eq:correlation-matrix}
\Sigma_{P_{\mathrm{Ly\alpha}}} = \frac{1}{N_{\text{mocks}}} \sum_{i=1}^{N_{\text{mocks}}} (P_{\mathrm{Ly\alpha}, i} - \langle P_{\mathrm{Ly\alpha}} \rangle)(P_{\mathrm{Ly\alpha}, i} - \langle P_{\mathrm{Ly\alpha}} \rangle)^\top
\end{equation}
To better visualize each covariance matrix, we define a corresponding correlation matrix $C$, which expresses the covariances between $i$-th and $j$-th bins in units of the corresponding $i$-th and $j$-th diagonal elements of the covariance matrix. A given $C_{ij}$ element in the correlation matrix is calculated with:
\begin{equation}
C_{ij} = \frac{\Sigma_{ij}}{\sqrt{\Sigma_{ii} \Sigma_{jj}}}
\end{equation}
Figure \ref{fig:covar-examples} shows two examples of covariance matrices, with parameters $z_{\text{mid}}$ = 6.4, $\Delta z$ = 1.0, $A_{z}$ = 12.4, $\Delta T_{\text{re}}$ = 29000, $\langle F \rangle$ = 0.1496, and  $z_{\text{mid}}$ = 8.7, $\Delta z$ = 16.0, $A_{z}$ = 11.5, $\Delta T_{\text{re}}$ = 36000, $\langle F \rangle$ = 0.1156. As can be seen, these covariance matrices present distinct off-diagonal values, especially towards wave numbers corresponding to smaller scales.

\subsection{Dataset of models at \texorpdfstring{$z=5.0$}{z=5.0}}\label{subsec:simulation-dataset}
Following the procedure outlined in this section, we created a dataset $\boldsymbol{\mathcal{D}}$ of 
models. To reduce the number of simulations that span the parameter space we wish to explore, 
we used Latin Hypercube Sampling (LHS) to create 750 different combinations of 
$\boldsymbol{\theta}_{\text{sim}} = \{z_{\text{mid}}, \Delta z, A_z, \Delta T_{\text{re}}\}$\footnote{
The $\langle F \rangle$ was excluded in this step, since it is added in post-processing, as explained in Subsection \ref{subsec:forward-modeling}}. Table \ref{tab:parameter-ranges} shows the ranges of values for our model parameters. These ranges were chosen to explore a parameter space that describes a wide variety of reionization histories. 

\begin{figure}
    \includegraphics[width=\columnwidth]{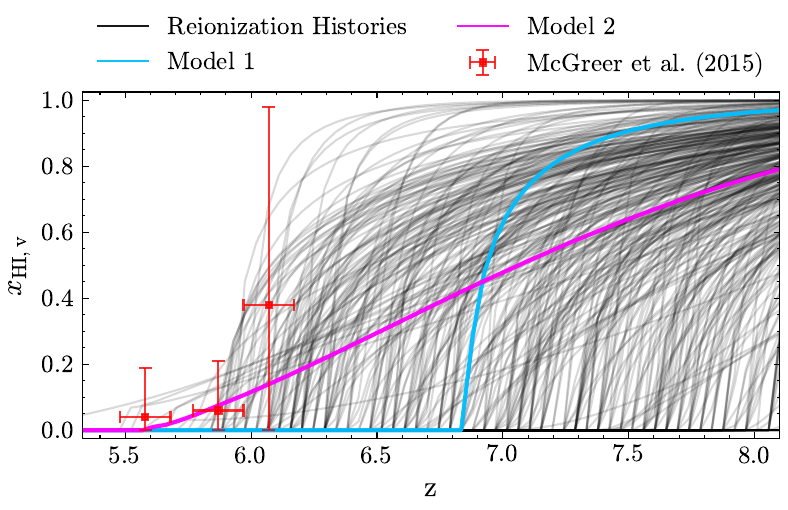}
    \caption{The 501 volume weighted reionization histories used to create our dataset of low resolution simulations (shown in black). These models were selected to be in agreement within $3\sigma$ with the measurements from \citet{McGreer2015} (shown in red). Highlighted in cyan and magenta are the reionization histories corresponding to Models 1 and 2, which are the to the same models as the top and bottom slices in Figure \ref{fig:simulation-slice}, with parameters $z_{\mathrm{mid}} = 7.1$, $\Delta z = 2.2$, $A_z = 8.3$, $\Delta T_{\mathrm{re}} = 9600$ K, and $z_{\mathrm{mid}} = 7.9$, $\Delta z = 4.5$, $A_z = 1.6$, $\Delta T_{\mathrm{re}} = 20000$ K respectively.}
    \label{fig:filtered_reion_histories}
\end{figure}

Additionally, to exclude 
unreasonable reionization histories that end too late, we filtered out all the parameter combinations that produced reionization histories that were not in agreement within $3 \sigma$ of the \textit{``dark pixels"} measurement from \citet{McGreer2015}. We selected this measurement because it is a robust and model-independent constraint at the tail end of the EoR. Since the measurements from \citet{McGreer2015} are volume-weighted neutral fractions, we converted the cell-based reionization redshift fields from AMBER into volume-weighted ionization histories. Specifically, we followed the procedure of \citet{Doughty2025}, computing the volume-weighted ionized fraction $x_{\text{HII,,v}}(z)$ as the fraction of total simulation volume in cells with $z_{\text{reion}} < z$. This left us with 501 combinations of $\boldsymbol{\theta}_{\text{sim}}$, as seen in Figure \ref{fig:filtered_reion_histories}. We executed Nyx simulations for all $\boldsymbol{\theta}_{\text{sim}}$, each with a simulation size of $L_{\text{box}} = 20$ $h^{-1}$ Mpc, and a resolution of $256^3$ voxels (as described in Section \ref{subsec:nyx-sims}). 

Since our objective is to test our method for parameter inference, we only output snapshots of these simulations at a redshift of $z=5.0$. Following the procedure outlined in Section \ref{subsec:forward-modeling}, $N_{\text{skwrs}}=5000$ Ly-$\alpha$ flux skewers were generated by introducing 9 different, regularly spaced values of $\langle F \rangle$ for each combination of $\boldsymbol{\theta}_{\text{sim}}$ (Table \ref{tab:parameter-ranges} lists the range of values $\langle F \rangle$). In total, we generated $4509$ models. Lastly, a mean power spectrum $\langle P_{\mathrm{Ly\alpha}}\rangle$, a set of $10^6$ mock observations $P_{\mathrm{Ly\alpha}, i}$ (though only the first 1000 are saved to reduce the memory requirements), and a covariance matrix $\Sigma_{P_{\mathrm{Ly\alpha}}}$ were generated for each model. In summary, we created a dataset of model-dependent $\boldsymbol{\mathcal{D}} = \{\langle P_{\mathrm{Ly\alpha}} \rangle, P_{\mathrm{Ly\alpha}, i}, \Sigma_{P_{\mathrm{Ly\alpha}}}\}$ as a function of our five model parameters $\boldsymbol{\theta} = \{z_{\text{mid}}, \Delta z, A_z, \Delta T_{\text{re}}, \langle F \rangle \}$. 

\begin{table}
	\centering
	\caption{Ranges of values for each of the $\theta_i$ parameters in our models. 
        These parameter ranges span a wide variety of reionization histories. }
	\label{tab:parameter-ranges}
	\begin{tabular}{lcc} 
		\hline
		$\theta_i$ & $\theta_{\text{min}}$ & $\theta_{\text{max}}$\\
		\hline
		$z_{\text{mid}}$     & 6.0    & 10.0    \\
		$\Delta z$           & 0.5    & 20.0    \\
            $A_{z}$              & 1.0    & 15.0    \\
            $\Delta T_{\text{re}}$      & 2000 K & 40000 K \\
		$\langle F \rangle$  & 0.1156 & 0.1700  \\
		\hline
	\end{tabular}
\end{table}

\section{Parameter Inference}\label{sec:parameter-inference}
To quantitatively constrain the parameters $\theta$ in our model, we use Bayesian inference, where:
\begin{equation}
    P(\theta|P_{\mathrm{Ly\alpha}, i}) = \frac{\mathcal{L}(P_{\mathrm{Ly\alpha}, i}|\theta) P(\theta)}{P(P_{\mathrm{Ly\alpha}, i})}
    \label{eq:bayes-theorem}
\end{equation}
The selection of reionization histories that we performed using the dark pixels measurement (see Section \ref{subsec:simulation-dataset}) imposed a prior on $\theta$, which we thus define to be a uniform distribution delimited by a convex hull that encloses the parameter space covered by our dataset of simulations. 
The choice of using a convex hull ensures that our uniform prior distribution only covers the region of the parameter space where our simulations have been evaluated. Additionally, we assume a multivariate Gaussian likelihood, such that:
\begin{equation}
    \mathcal{L} = \frac{1}{\sqrt{\text{det}(\Sigma_{P})(2\pi)^n }} \text{exp}\bigg( -\frac{(P_{i} - \langle P \rangle)^\top \Sigma_{P}^{-1}  (P_{i} - \langle P \rangle)}{2} \bigg)
	\label{eq:likelihood}
\end{equation}
Where $n$ is the number of $k$ bins. Here we have dropped the $Ly\alpha$ subscript from $\langle P_{\mathrm{Ly\alpha}} \rangle$, $P_{\mathrm{Ly\alpha}, i}$, and $\Sigma_{P_{\mathrm{Ly\alpha}}}$ for readability. Given that our dataset only provides us with a limited number of model evaluations within the parameter space that we are exploring, we construct emulators for $\langle P_{\mathrm{Ly\alpha}} \rangle$ and $\Sigma_{P_{\mathrm{Ly\alpha}}}$, which are described in Section \ref{subsec:emulators}. In addition, we opt to use Hamiltonian Monte Carlo to accelerate the inference procedure, described in Section \ref{subsec:hmc}.

\subsection{Neural Emulators}\label{subsec:emulators}
The objective of our neural emulators is to allow us to approximate the model evaluation at any point in the parameter space defined by the prior distribution, despite only having a relatively small number of simulations in our dataset $\mathcal{D}$. The basic idea of a neural emulator is to train a neural network such that it takes $\boldsymbol\theta$ as input, and outputs some model-dependent quantity. In our model, both $\langle P_{\mathrm{Ly\alpha}} \rangle$ and $\Sigma_{P_{\mathrm{Ly\alpha}}}$ have a dependence on $\boldsymbol\theta$. Thus, we build two separate neural emulators.

\begin{figure*}
    \centering
	\includegraphics[width=1.9\columnwidth]{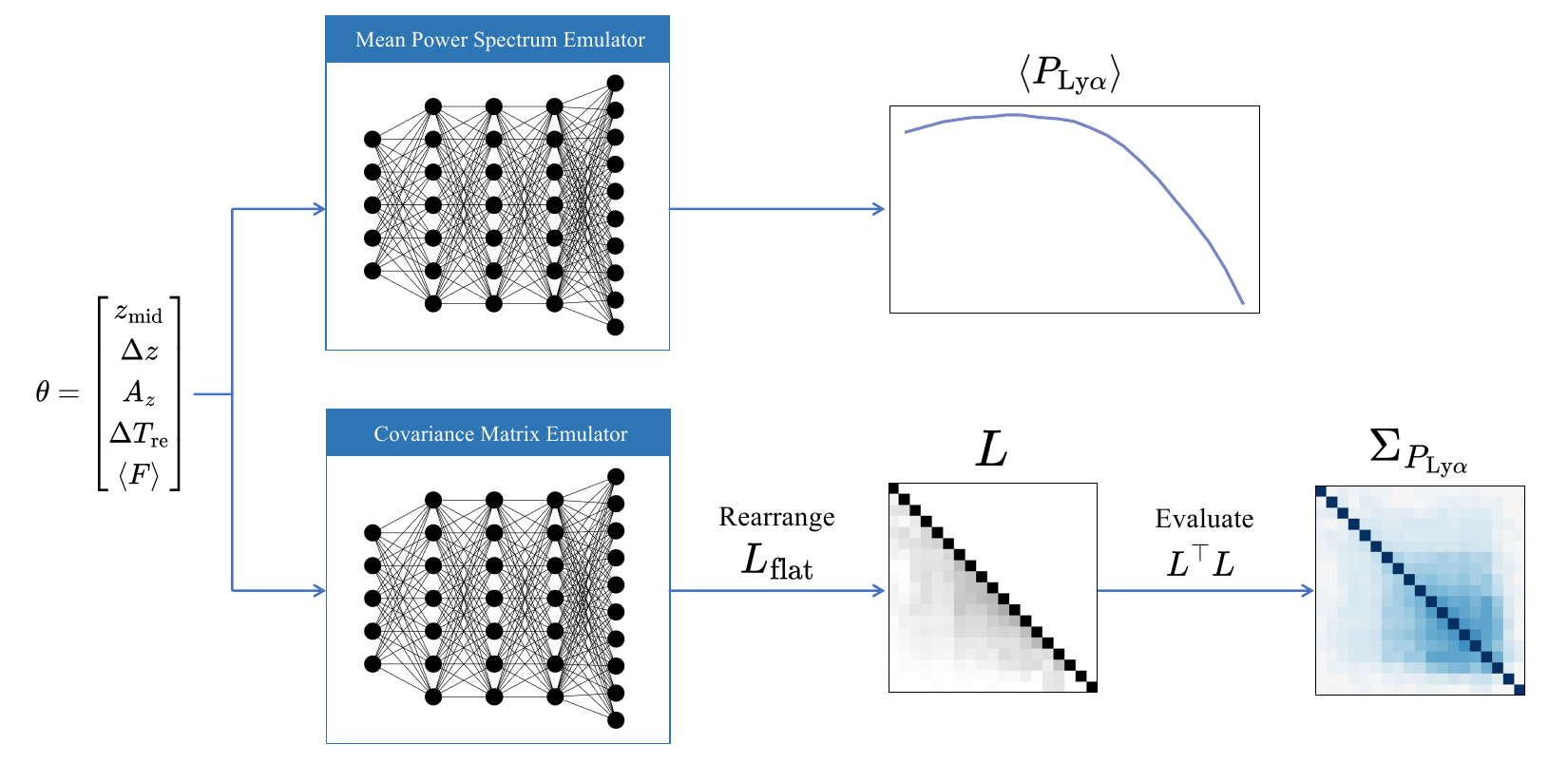}
    \caption{Diagram of the emulators. The top portion shows the emulator for the mean power spectrum, where a fully connected, feed-forward neural network is used to emulate the power spectrum directly from the input parameters $\theta$. The bottom of the diagram shows the emulator for the covariance matrix. In this case, a neural network predicts $L_{\text{flat}}$ from $\theta$. $L_{\text{flat}}$ is then rearranged into $L$, from which an emulated covariance matrix can be obtained.}
    \label{fig:emulators-diagram}
\end{figure*}

\subsubsection{Power Spectrum Emulator}\label{subsec:mean-emulator}
To emulate $\langle P_{\mathrm{Ly\alpha}} \rangle$, we employ a fully connected, feed-forward neural network. This neural network is trained to output $\log_{10}\langle P_{\mathrm{Ly\alpha}} \rangle$ as a function of the five parameters in our model $\boldsymbol\theta$. We adopt the standard Mean Squared Error (MSE) loss for training, and both the input $\boldsymbol\theta$ and the output $\log_{10}\langle P_{\mathrm{Ly\alpha}} \rangle$ are normalized using a MinMax scaler. The top section of Figure \ref{fig:emulators-diagram} shows a diagram of the emulation process for $\langle P_{\mathrm{Ly\alpha}} \rangle$.

\subsubsection{Covariance Matrix Emulator}\label{subsec:covar-emulator}
Careful consideration is required to emulate the covariance matrices $\Sigma_{P_{\mathrm{Ly\alpha}}}$. Following the same approach as in \citet{Kang2024}, and to ensure that the emulated $\Sigma_{P_{\mathrm{Ly\alpha}}}$ are symmetric and positive definite, we take advantage of Cholesky decomposition:
\begin{equation}
    A = LL^*
	\label{eq:cholesky}
\end{equation}
where $A$ is a Hermitian positive-definite matrix and $L$ is a lower triangular matrix known as the Cholesky factor. Since the covariance matrices in our model are real valued, we can use Equation \ref{eq:cholesky} to find the Cholesky factor for each covariance matrix such that $\Sigma_{P_{\mathrm{Ly\alpha}}}=LL^\top$. In addition, we reorder each $L$ into a 1D array $L_{\text{flat}}$ where the first elements correspond to the diagonal values of $L$ and the rest of the elements correspond to the off-diagonal values. This allows us to use a separate fully connected, feed-forward neural network to emulate $\Sigma_{P_{\mathrm{Ly\alpha}}}$ as a function of $\boldsymbol\theta$. Lastly, to ensure that the diagonal elements of the emulated $L$ are positive, we apply an exponential activation function for these elements only, as done in \citet{Kang2024}. For training, we adopt the MSE loss calculated on $L_{\text{flat}}$, and use a MinMax scaler for the input and output of this neural network. The lower portion of Figure \ref{fig:emulators-diagram} shows a diagram of the emulation process for $\Sigma_{P_{\mathrm{Ly\alpha}}}$.

\subsection{Hamiltonian Monte Carlo}\label{subsec:hmc}
A popular choice to accelerate parameter inference is Hamiltonian Monte Carlo (HMC), a gradient-based modification of the Metropolis-Hastings sampling algorithm. HMC improves the efficiency of parameter space exploration by proposing samples that have a higher probability of being accepted \citep{Duane1987, Neal2011}. To achieve this, HMC formulates the problem by creating an analogy between the exploration of the parameter space and the dynamic evolution of a particle living in a phase space defined by the particle's position (chosen to be equal to $\boldsymbol{\theta}$) and momentum $\boldsymbol{p}$. Typically, this particle is assigned the following Hamiltonian $\mathcal{H(\boldsymbol{\theta}, \boldsymbol{p})}$:
\begin{equation}
    \mathcal{H(\boldsymbol{\theta}, \boldsymbol{p})} = \frac{1}{2} \boldsymbol{p}^{T} \mathcal{M}^{-1} \boldsymbol{p} + \boldsymbol{U}(\boldsymbol{\theta})
	\label{eq:hamiltonian}
\end{equation}
Where $\mathcal{M}$ is known as the mass matrix, and $\boldsymbol{U}(\boldsymbol{\theta})$ is the potential energy of the particle. The choice of $\boldsymbol{U}(\boldsymbol{\theta})$ can depend on the specific application, but the following is generally used:
\begin{equation}
    \boldsymbol{U}(\boldsymbol{\theta}) = -ln\big(\mathcal{L}(\boldsymbol{d}|\boldsymbol{\theta}) P(\boldsymbol{\theta}) \big)
	\label{eq:potential}
\end{equation}
Intuitively, this choice for $\boldsymbol{U}(\boldsymbol{\theta})$ causes the regions of high density in the posterior $P(\boldsymbol{\theta}|\boldsymbol{d})$ to be located in the regions with low potential energy $\boldsymbol{U}(\boldsymbol{\theta})$ (see Equation~\ref{eq:bayes-theorem}). HMC takes advantage of this and increases the efficiency of the standard Metropolis-Hastings algorithm by proposing samples that correspond to low energy levels of $\mathcal{H(\boldsymbol{\theta}, \boldsymbol{p})}$, which have a higher chance of being accepted. Therefore, the momentum $\boldsymbol{p}$ is randomly sampled from a multivariate Gaussian distribution with zero mean and a covariance matrix given by $\mathcal{M}$ at the beginning of every step of the chain, forcing the hypothetical particle to spend more time exploring the parameter space in regions of low $\boldsymbol{U}(\boldsymbol{\theta})$. For a target distribution with $D$ dimensions, the cost of an independent sample produced by HMC scales as $O(D^{5/4})$, a significant improvement over the scaling of $O(D^{2})$ of the standard Metropolis-Hastings algorithm \citep{Neal2011}.

Unfortunately, adopting HMC in most problems is not possible. This is because the numerical integration of $\mathcal{H(\boldsymbol{\theta}, \boldsymbol{p})}$ performed for every new proposed sample in the HMC algorithm requires $\mathcal{L}(\boldsymbol{d}|\boldsymbol{\theta})$ (and in turn, the model evaluation) to be fully differentiable, which is typically not the case in most inference problems. However, using neural networks as surrogates for our model completely solves this problem, because we are able to calculate their derivatives with respect to $\boldsymbol\theta$ via auto-differentiation and back propagation \citep{Rumelhart1986}. This allows us to fully take advantage of HMC. 

\subsection{Implementation with  JAX}\label{subsec:jax}
In order to easily apply the principles of our approach to perform inference, we have developed a pipeline that partially automates some of the required steps. To facilitate the auto-differentiation evaluations of the neural emulators, we opt to use \texttt{JAX} \citep{Bradbury2018}. \texttt{JAX} is a high-performance Python package mainly designed to work as a machine learning framework, and it has been used to create auto-differentiable code in a variety of studies \citep[e.g.][]{Piras2023, Ataei2024, Citrin2024}. In \texttt{JAX}, Python functions are traced into functional and stateless representations, on which it applies predetermined transformations to perform automatic differentiation. This approach makes JAX more efficient than other machine learning frameworks, and is the reason we decided to build our pipeline around it. Our pipeline is built to perform the following:

\begin{enumerate}
    \item \textbf{Training the emulators:} Given a dataset of $\{\boldsymbol{\theta}, \langle P_{\mathrm{Ly\alpha}} \rangle, \Sigma_{P_{\mathrm{Ly\alpha}}}\}$, the pipeline trains two separate emulators for $\langle P_{\mathrm{Ly\alpha}} \rangle$ and $\Sigma_{P_{\mathrm{Ly\alpha}}}$. For training, we adopt a cosine decay schedule for the learning rate, and interrupt training if there is no improvement over 150 consecutive epochs. We use \texttt{Flax} to construct and train the neural networks \citep{Heek2024}. \vspace{0.2cm}

    \item \textbf{Hyper-parameter tuning:} The pipeline is able to perform the hyper-parameter tuning for both emulators. We use \texttt{Optuna} to carry out this procedure \citep{Akiba2019}. \vspace{0.2cm}
    
    \item \textbf{Parameter inference:} Given a set of observations $\{P_{\mathrm{Ly\alpha}, i}\}$, the pipeline performs parameter inference for $\boldsymbol{\theta}$ using the trained emulators. To apply HMC, we use the implementation of the No-U-Turn algorithm \citep{Hoffman2011} in the \texttt{Numpyro} package \citep{Bingham2019}.
\end{enumerate}

A version of this pipeline that can be applied to other domain-specific problems has been made publicly available in the Differential Neural Emulators with Hamiltonian Monte Carlo (\texttt{DNE + HMC}) GitHub repository\footnote{\href{https://github.com/enigma-igm/LAF_emulator}{\faGithub \texttt{ DNE + HMC}}}.

\section{Results}\label{sec:results}
We split our dataset $\boldsymbol{\mathcal{D}}$ into $70\%$ - $10\%$ - $20\%$ partitions for the training, validation, and test sets respectively. To ensure the robustness of our emulators, we performed this split on the original 501 simulations, instead of the final 4509 models that resulted after adding the mean flux $\langle F \rangle$. We then applied our pipeline to train the two emulators and perform parameter inference on a set of mock observations. To perform the hyper-parameter tuning of both emulators, we opt to vary the number of hidden layers $n_h$, the number of units for each layer $n_{u,i}$, the initial learning rate $l_{r}$, and the number of training epochs $n_e$. Table \ref{tab:hparam-results} in Appendix \ref{app:hparam-tuning} shows the results of the hyper-parameter tuning.

\subsection{Emulator Performance}\label{subsec:emulator-performance}
The MSE loss that we employed during the training of both emulators lacks interpretability. Consequently, we employ more interpretable metrics to verify the performance of both emulators, and calculate them using the test set exclusively.

To evaluate the performance of the $\langle P_{\mathrm{Ly\alpha}} \rangle$ emulator, we use the mean absolute percentage error (MAPE) as our main metric. For a single power spectrum, the MAPE is calculated by:
\begin{equation}
\text{MAPE}{\hspace{0.04cm}i} = \frac{1}{n{b}}\sum_{i=1}^{n_b}\frac{|\langle\hat{P}{\mathrm{Ly\alpha}}\rangle_i - \langle P{\mathrm{Ly\alpha}} \rangle_i|}{\langle P_{\mathrm{Ly\alpha}} \rangle_i}
\label{eq:mape}
\end{equation}
Where the subscript corresponds to the $i$-th $k$ mode, $\langle\hat{P}_{\mathrm{Ly\alpha}}\rangle$ is the emulated power spectrum, and $n_b$ is the number of $k$ bins. With this definition, MAPE provides an interpretable estimate of the relative error that our emulator incurs.

To verify if the emulation error is sufficiently low to correctly approximate $\langle P_{\mathrm{Ly\alpha}} \rangle$ as a function of our model parameters, we compare them to the expected level of observational uncertainty that is expected (in our case, this is defined by our assumptions in the forward modeling, see Section~\ref{subsec:forward-modeling}). Specifically, we compute the square root of the diagonal elements of the model-dependent covariance matrices, $\sqrt{(\Sigma_{P_{\mathrm{Ly\alpha}}})_{ii}}$, which correspond to the observational 1$\sigma$ error bars for each $k$ bin. This comparison is critical, as the required accuracy of the emulator is ultimately determined by the signal-to-noise (S/N) ratio of the data. Ideally, the emulation error should be smaller than the observational uncertainties, and will have a negligible effect on parameter inference. 

\begin{figure}
    \includegraphics[width=\columnwidth]{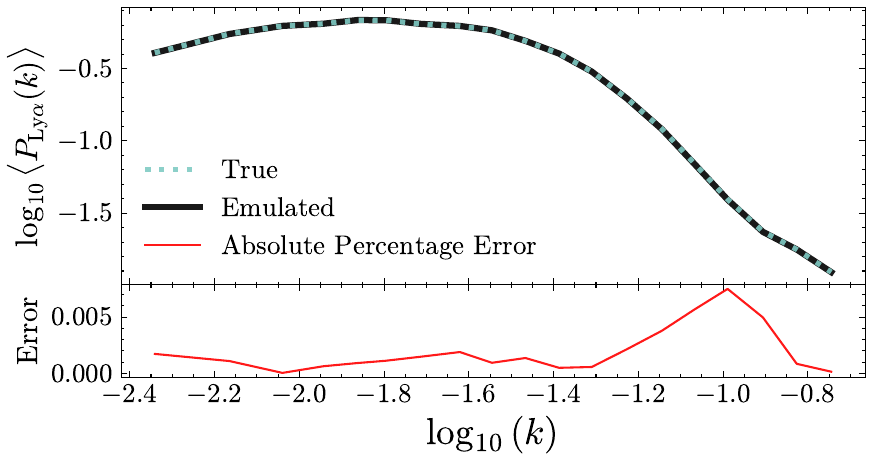}
    \caption{Example of an emulated power spectrum in the test dataset, with an error in the 50th percentile. The emulated power spectrum is shown in the solid black line, while the true power spectrum is shown in the light blue dashed line. The bottom panel shows the absolute relative error as a function of $k$. }
    \label{fig:mean-emulation-example}
\end{figure}

\begin{figure}
    \includegraphics[width=\columnwidth]{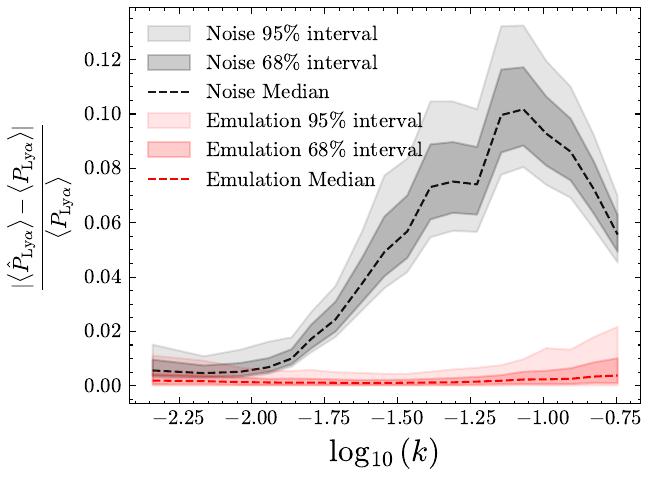}
    \caption{Distribution of the power spectrum emulator's relative errors as a function of $k$, evaluated on the test set. The dashed red line represents the median emulator error across all test models, while the dark and light red shaded regions indicate the central 68\% and 95\% intervals, respectively. For comparison, the dashed black line shows the median of the expected 1$\sigma$ observational uncertainty, computed as the square root of the diagonal elements of the model-dependent covariance matrices. Across all scales, the emulator’s typical error remains below the expected noise level.}
    \label{fig:power-spectrum-performance}
\end{figure}

Verifying the performance of our $\langle P_{\mathrm{Ly\alpha}} \rangle$ emulator with the full test set, we obtain a median MAPE of 0.15\%, indicating very high accuracy. Figure \ref{fig:mean-emulation-example} shows an example at the 50th percentile of emulator error, with the relative error as a function of $k$ displayed in the bottom panel. Similar trends in relative error appear throughout the test dataset. Figure \ref{fig:power-spectrum-performance} summarizes the absolute percentage error distribution as a function of $k$, with the solid dashed red line indicating the median and shaded regions showing the central 68\% and 95\% intervals. Across all $k$ modes, most of errors remain at the sub-percent level, Even in extreme cases, 95\% of test examples maintain errors below 2\%, confirming that even the worst-case emulation performance remains low. This distribution of relative errors remains well below the expected observational noise level, given the assumptions in our forward modeling. This indicates that the emulator performs significantly better than necessary for current datasets, even at high-$k$, where both the emulator error and observational uncertainties increase due to decreased S/N.

\subsubsection{Performance of Covariance Matrix Emulator}\label{subsec:performance-covar-emulator}

\begin{figure*}
    \centering
	\includegraphics[width=1.98\columnwidth]{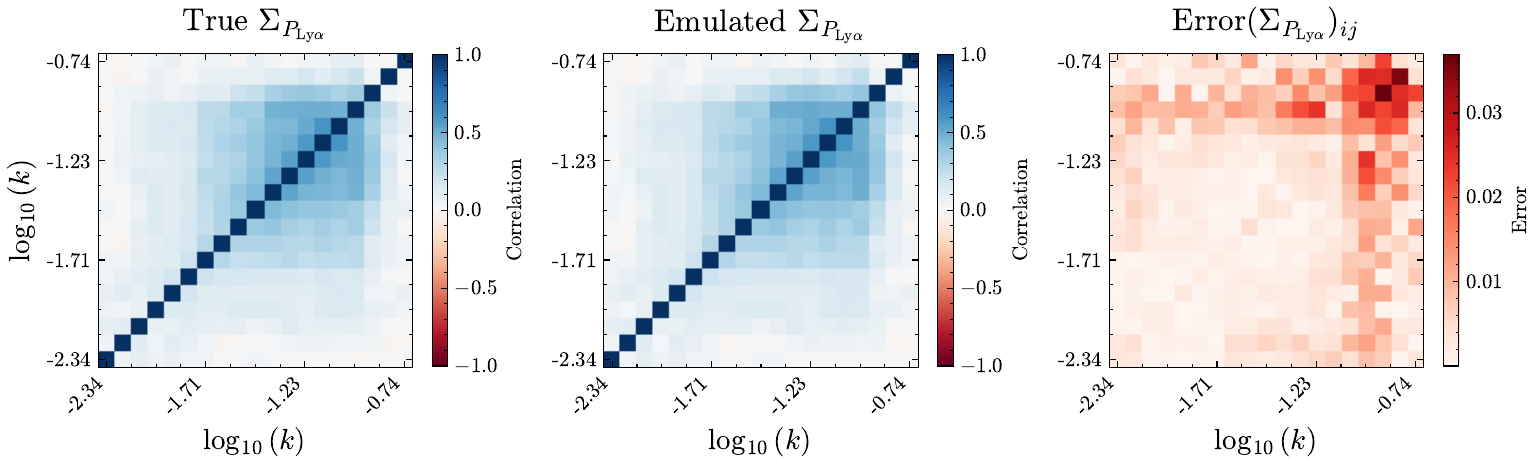}
    \caption{Example of an emulated covariance matrix in the test dataset, with an error in the 50th percentile. The true covariance matrix is shown in the left, while the emulated covariance matrix is shown in the middle. Both covariance matrices are shown as correlation matrices as described by Equation \ref{eq:correlation-matrix}. The heat map on the right shows the error as defined by Equation \ref{eq:covar_error} for this example.}
    \label{fig:covar-emulation-example}
\end{figure*}

\begin{figure}
    \includegraphics[width=0.95\columnwidth]{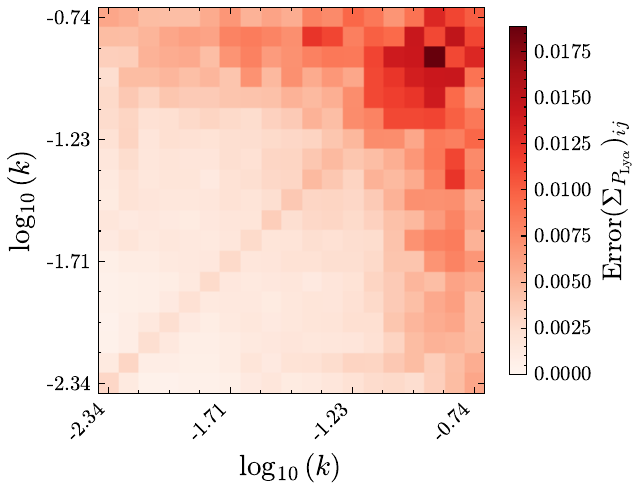}
    \caption{Median of the $\Sigma_{p_{\mathrm{Ly\alpha}}}$ emulator's errors as defined by Equation \ref{eq:covar_error}, calculated with the test set. As observed, the emulator incurs in a larger error for elements in the covariance matrix that correspond to the small scales.}
    \label{fig:covar-performance}
\end{figure}

To evaluate the performance of the $\Sigma_{P_{\mathrm{Ly\alpha}}}$ emulator, we define the error of the $ij$-th element of a single covariance matrix with the following:
\begin{equation}
    \text{Error}(\Sigma_{P_{\mathrm{Ly\alpha}}})_{\hspace{0.04cm}ij} = \frac{|\hat{\Sigma}_{P_{\mathrm{Ly\alpha}},ij}-\Sigma_{P_{\mathrm{Ly\alpha}},ij}|}{\sqrt{\Sigma_{P_{\mathrm{Ly\alpha}},ii}\Sigma_{P_{\mathrm{Ly\alpha}},jj}}}
	\label{eq:covar_error}
\end{equation}
Where $\hat{\Sigma}_{P_{\mathrm{Ly\alpha}}}$ is the emulated covariance matrix. Equation \ref{eq:covar_error} represents the emulation error relative to the product of the corresponding $i$-th and $j$-th diagonal elements of the true covariance matrix. Figure \ref{fig:covar-emulation-example} shows an example of an emulated covariance matrix in the test dataset, with an error in the 50th percentile, as calculated by taking the average of Equation \ref{eq:covar_error} over the matrix elements. 
Additionally, Figure \ref{fig:covar-performance} shows the median (for all matrix elements)
of this metric calculated over the test dataset using our covariance matrix emulator. Additionally, setting $i=j$ in Equation \ref{eq:covar_error} allows us to simplify it to:

\begin{equation}
    \text{Error}(\text{diag}(\Sigma_{P_{\mathrm{Ly\alpha}}}))_{\hspace{0.01cm}i} = \frac{|\hat\sigma_i - \sigma_i|}{\sigma_i}
	\label{eq:covar_diag_error}
\end{equation}

\begin{figure}
    \includegraphics[width=\columnwidth]{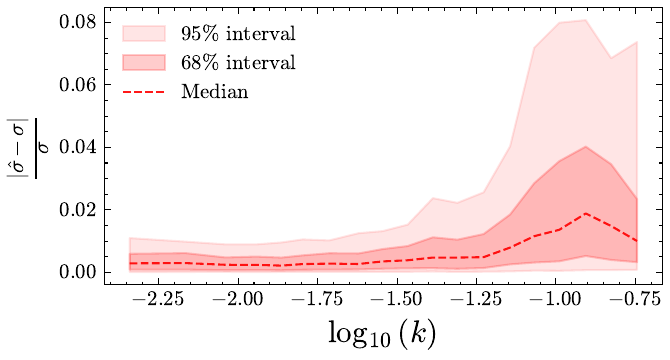}
    \caption{Distribution of the $\Sigma_{p_{\mathrm{Ly\alpha}}}$ emulator's errors on the diagonal of the covariance matrices as a function of $k$, calculated with the test set. The black solid line represents the median of this quantity, while the dark red and light red shaded regions contain the regions with 68\% and 95\% of all errors. This figure shows that the covariance matrix emulator incurs in a sub-5\% error across all scales in the majority of cases.}
    \label{fig:covar-diag-performance}
\end{figure}

Where $\hat\sigma_i$ and $\sigma_i$ are the $i$-th diagonal element of the emulated and true covariance matrices respectively. Similarly to Equation \ref{eq:mape}, this simplification is the MAPE calculated on the diagonal elements of the covariance matrices, allowing us to interpret the performance of our covariance matrix emulator. Our covariance matrix emulator obtains a median of 0.0042 for the metric given by Equation~\ref{eq:covar_error}, calculated with the entire test set. Figure \ref{fig:covar-diag-performance} shows the distribution of this metric calculated with the test dataset as a function of $k$. The solid black line shows the median of this error as a function of the diagonal element corresponding to a $k$ bin, while the shaded regions show the intervals that contain 68\% and 95\% of all errors. As we can observe, the emulator produces a higher error at smaller scales, but consistently produces sub 5\% error across all scales.

\subsection{Parameter Estimation}\label{subsec:parameter-estimation}

\begin{figure*}
    \centering
    \includegraphics[width=0.95\columnwidth]{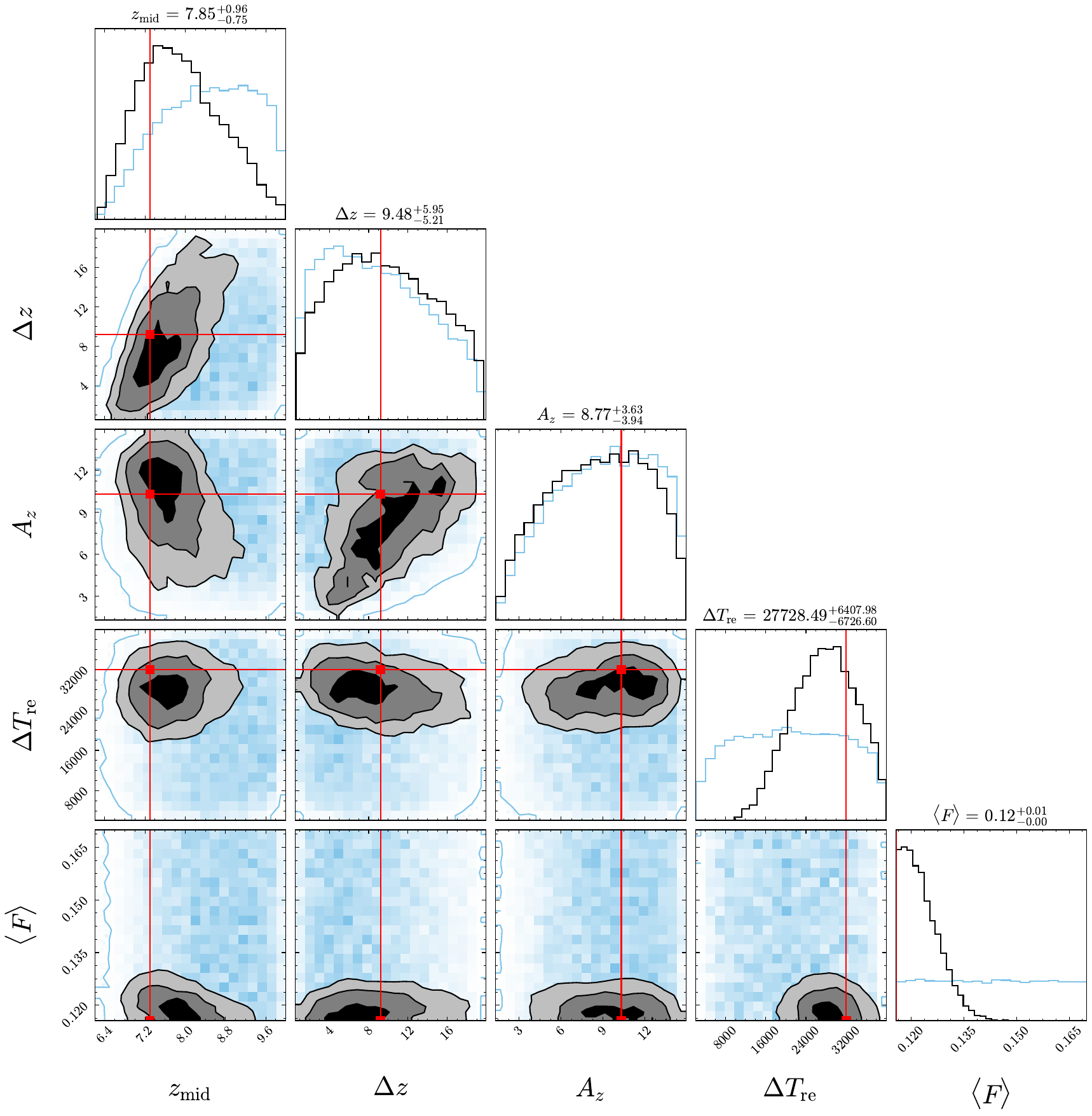}
    \includegraphics[width=0.95\columnwidth]{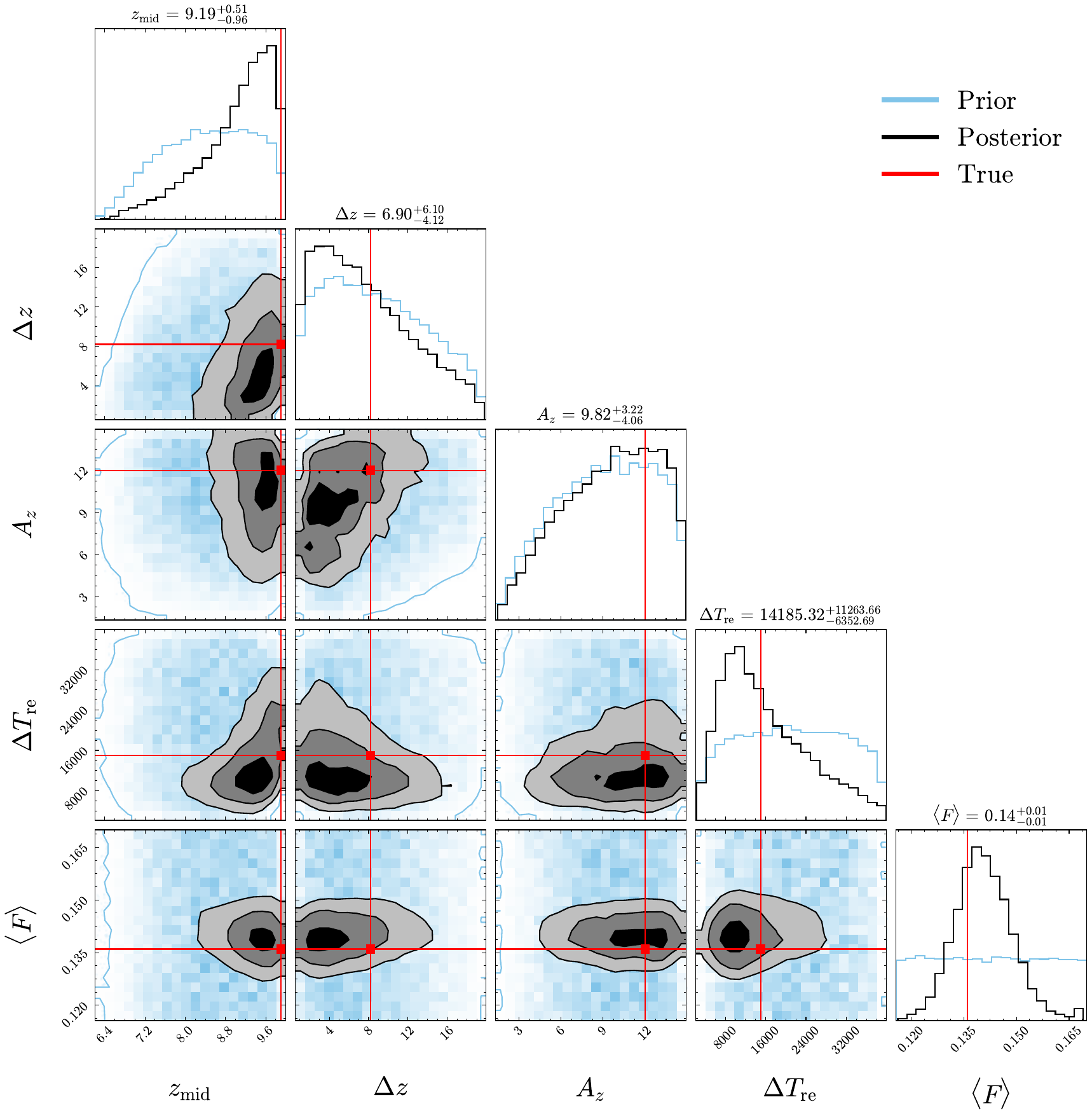}
    \caption{The posterior distributions obtained for two randomly selected mock observations. The corner plots show the posterior distributions (in black) overlaid on top of the priors (in light blue). The true parameters of each mock observation are plotted in red. The median of the marginalized posteriors for each parameter is shown on top of the 1D histograms, along with the uncertainty of their estimation (given by the $16$-th and $84$-th quantiles). The true parameters for the corner plot on the left are $z_{mid}$ = 7.3, $\Delta z$ = 9.2, $A_{z}$ = 10.3, $\Delta T_{\text{re}}$ = 32000 K, $\langle F \rangle$ = 0.1156, while the true parameters for the mock on the right are $z_{mid}$ = 9.9, $\Delta z$ = 8.2, $A_{z}$ = 12.0, $\Delta T_{\text{re}}$ = 15000 K, $\langle F \rangle$ = 0.1360.}
    \label{fig:corner-plots}
\end{figure*}

\begin{figure*}
    \centering
    \includegraphics[width=0.97\columnwidth]{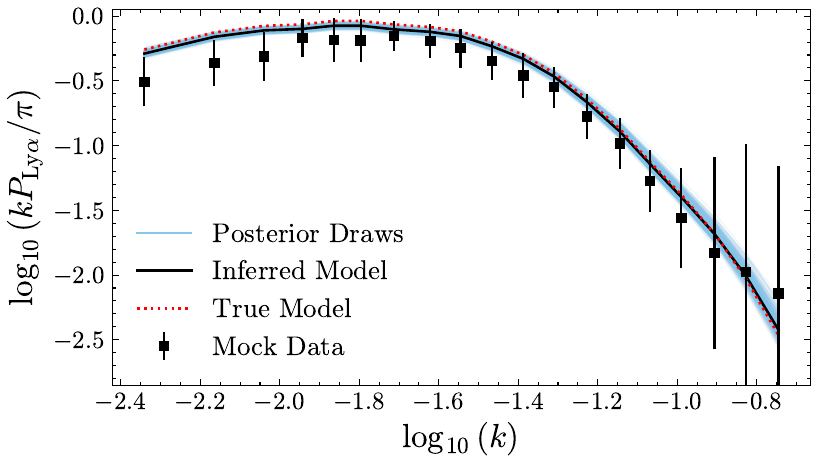}
    \includegraphics[width=0.97\columnwidth]{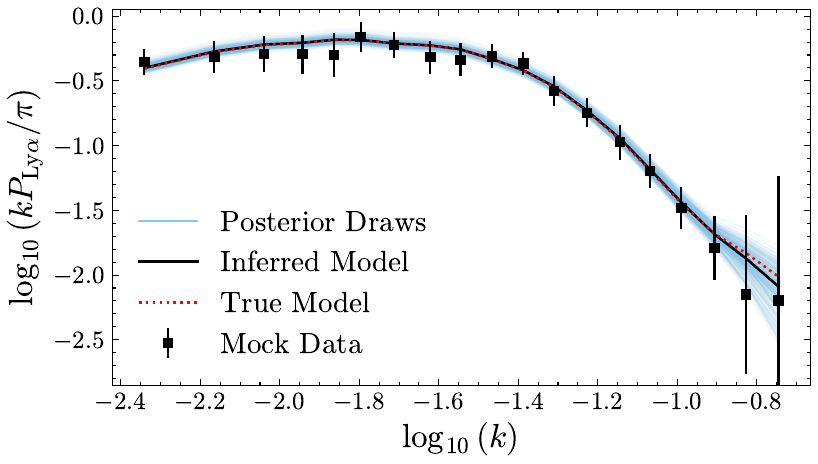}
    \caption{The inferred model plots for both posterior distributions in Figure \ref{fig:corner-plots}. For comparison, we show the true model (red dotted line), and 500 randomly selected posterior draws. We also plot the mock observations (square markers) along with their errors, which are taken to be equal to the square root of the diagonal elements of the covariance matrix corresponding to the true model respectively.}
    \label{fig:inferred-models}
\end{figure*}

To test the inference method in our approach, we apply our pipeline to perform parameter estimation on a subset of 100 randomly selected mock observations. For each parameter estimation, we set HMC to have 4 chains, where each chain takes 1000 steps for warm-up (i.e. ``burn-in''), and 2000 extra steps for sampling. 
Figure \ref{fig:corner-plots} shows the posterior distributions obtained for two separate mock observations, corresponding to models with true parameters $z_{\text{mid}}$ = 7.3, 
$\Delta z$ = 9.2, $A_{z}$ = 10.3, $\Delta T_{\text{re}}$ = 32000 K, $\langle F \rangle$ = 0.1156 and $z_{mid}$ = 9.9, $\Delta z$ = 8.2, $A_{z}$ = 12.0, $\Delta T_{\text{re}}$ = 15000 K, $\langle F \rangle$ = 0.1360 respectively, which are colored red. These corner plots also show the prior distributions for each parameter (defined by a convex hull, as described in Section \ref{sec:parameter-inference}).

To be able to compare the true underlying model (i.e., the true $\langle P_{\mathrm{Ly\alpha}} \rangle$) against the results of each inference, we use our mean power spectrum emulator to generate all the individual $\langle P_{\mathrm{Ly\alpha}} \rangle$ corresponding to every sample in a given posterior distribution. We then obtain the inferred model by calculating the median of all generated power spectra for each $k$ bin. Figure \ref{fig:inferred-models} shows the plots of the inferred models obtained following this procedure for the two examples of Figure \ref{fig:corner-plots}. This figure also shows the mock data, and a set of 500 randomly chosen models from the posterior distribution.

\subsubsection{Inference Test}\label{subsec:inference-test}
\begin{figure}
    \includegraphics[width=0.9\columnwidth]{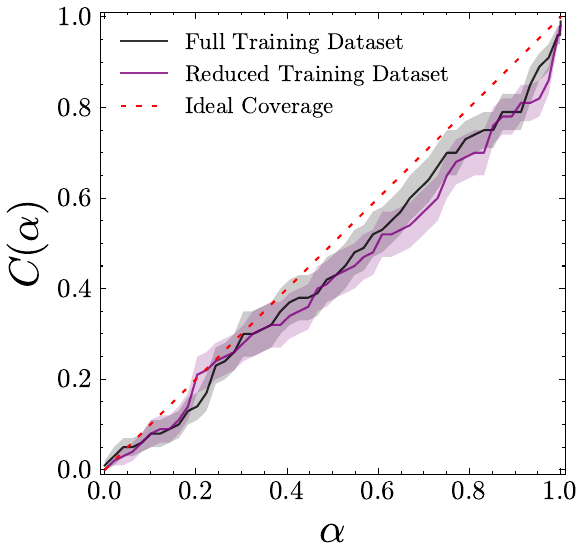}
    \caption{Coverage plot that we obtained with the results on performing the inference test on the same 100 randomly selected mock observations of the power spectrum using the emulators trained on the full training dataset (black) and the reduced training dataset (purple). The shaded regions show the Poisson errors of both inference tests.}
    \label{fig:coverage-plot}
\end{figure}

To properly characterize the robustness and quality of the posteriors that we obtain with our method, we decided to construct a coverage plot as an inference test. Provided that you have a posterior distribution, the coverage probability $C(\alpha)$ for a credibility level $\alpha$ is defined as the probability of the true value of the parameters being within the volume that encloses the corresponding credibility contour. The complete formalism of the method that we use to estimate the coverage probability from our set of 100 posterior distributions is detailed in \citet{Hennawi2025}, 
where the authors used the same inference test to assess the reliability of their quasar continuum reconstruction inference method. In order to satisfy this inference test, the estimated coverage probability $C(\alpha)$ should be equal to the posterior credibility for all levels. Figure \ref{fig:coverage-plot} shows the coverage probability that we get from our set of 100 posterior distributions. As we can see, the estimated coverage that we obtain lies marginally below the ideal line. This means that our posterior distributions are either slightly overconfident or biased. A detailed discussion on the interpretation of this inference test can be found in Section \ref{subsec:inference-test-discussion}.

\subsection{Reducing the size of the dataset}\label{subsec:reduced-dataset}
As explained in Section \ref{subsec:simulation-dataset}, the complete dataset of model dependent power spectra and covariance matrices was generated from 501 different low resolution simulations. In practice, if we were to create a dataset from high resolution simulations \citep[i.e with $\sim2048^3$ voxels per simulation cube required to properly resolve small scales, see][]{Doughty2023a}, 
we would not expect to be capable of performing hundreds of simulations, as this is expensive in terms of both computational resources and storage requirements. Consequently, we chose to test how the performance of our method would deteriorate with a significantly smaller number of simulations in our training and validation sets. 

\begin{figure}
    \includegraphics[width=\columnwidth]{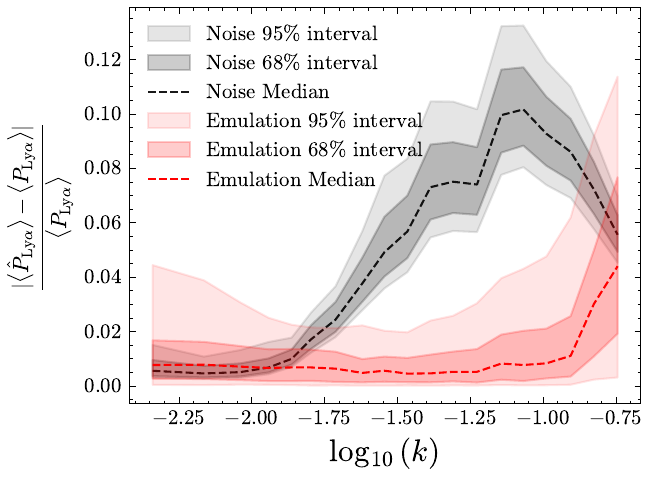}
    \caption{Distribution of the power spectrum emulator's errors as a function of $k$, calculated on the test set when trained with only 100 simulations. The black solid line shows the median of the absolute percentage error, while the dark red and light red shaded regions indicate the 68\% and 95\% intervals, respectively. The dashed black line represents the median observational uncertainty, estimated from the square root of the diagonal of the covariance matrix across the dataset. While the emulator still achieves sub-5\% accuracy across most $k$ bins, the increase in error—especially at small and large scales—makes it comparable to the expected observational noise in those regimes.}
    \label{fig:power-spectrum-performance-reduced}
\end{figure}

To do a direct comparison against the full dataset, we kept the test set exactly the same, and randomly selected 100 simulations to serve as the reduced dataset. These were split into 70 and 30 simulations for the training and validation sets respectively. We then applied our pipeline to train two new emulators with the reduced dataset, and performed hyper-parameter tuning following the same procedure as described in Section \ref{subsec:emulator-performance}. The mean emulator obtained has a MAPE of 0.012 on the test dataset compared to 0.0015 obtained with the full dataset. Figure \ref{fig:mean-emulation-example-reduced} (see Appendix \ref{app:reduced-performance}) shows an example of an emulated power spectrum with an error in the 50th percentile in the test dataset, while the distribution of this emulator's error as a function of $k$ is shown in Figure \ref{fig:power-spectrum-performance-reduced}. As we can appreciate, there is a substantial increase in emulation error, especially at smaller scales. Notably, the emulator's error becomes comparable to the expected observational uncertainty at both ends of $k$ values in our dataset. This illustrates the practical limit of how few simulations can be used before emulation performance starts to significantly degrade the scientific utility of the predictions.


\begin{figure}
    \includegraphics[width=\columnwidth]{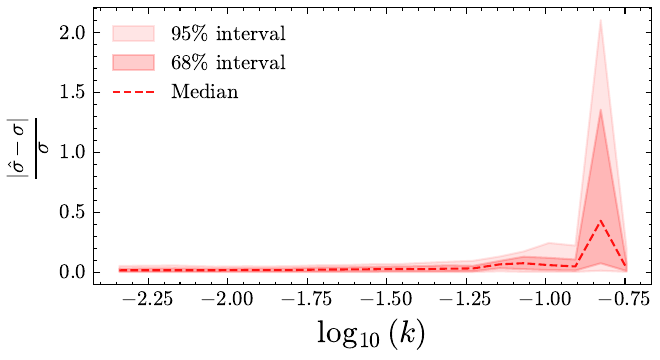}
    \caption{Distribution of the errors on the diagonal of the covariance matrices as a function of $k$ obtained with $\Sigma_{p_{\mathrm{Ly\alpha}}}$ emulator trained on the reduced dataset. The black solid line represents the median of this quantity, while the dark red and light red shaded regions contain the regions with 68\% and 95\% of all errors. This figure shows that the covariance matrix emulator incurs in a small error across most scales, but suffers of large errors at small scales.}
    \label{fig:covar-diag-performance-reduced}
\end{figure}

The new covariance matrix emulator trained on the reduced dataset obtained an average error of 0.031 for the metric given by Equation \ref{eq:covar_error} calculated on the test dataset, compared to 0.0042 for the full dataset. Figure \ref{fig:covar-emulation-example-reduced} in Appendix \ref{app:reduced-performance} shows an example of an emulated covariance matrix with an error in the 50th percentile in the test dataset, while Figure \ref{fig:covar-diag-performance-reduced} shows the distribution of relative errors in the diagonal elements also calculated with the test dataset. It is evident that reducing the training dataset dramatically affects the performance of this emulator, with an important increase in error at small scales. Although this is not optimal, the emulator still obtains a relatively small error on average. A more detailed discussion of the performance of these emulators can be found in Section \ref{subsec:emulator-performance-discussion}.

\begin{figure}
    \includegraphics[width=\columnwidth]{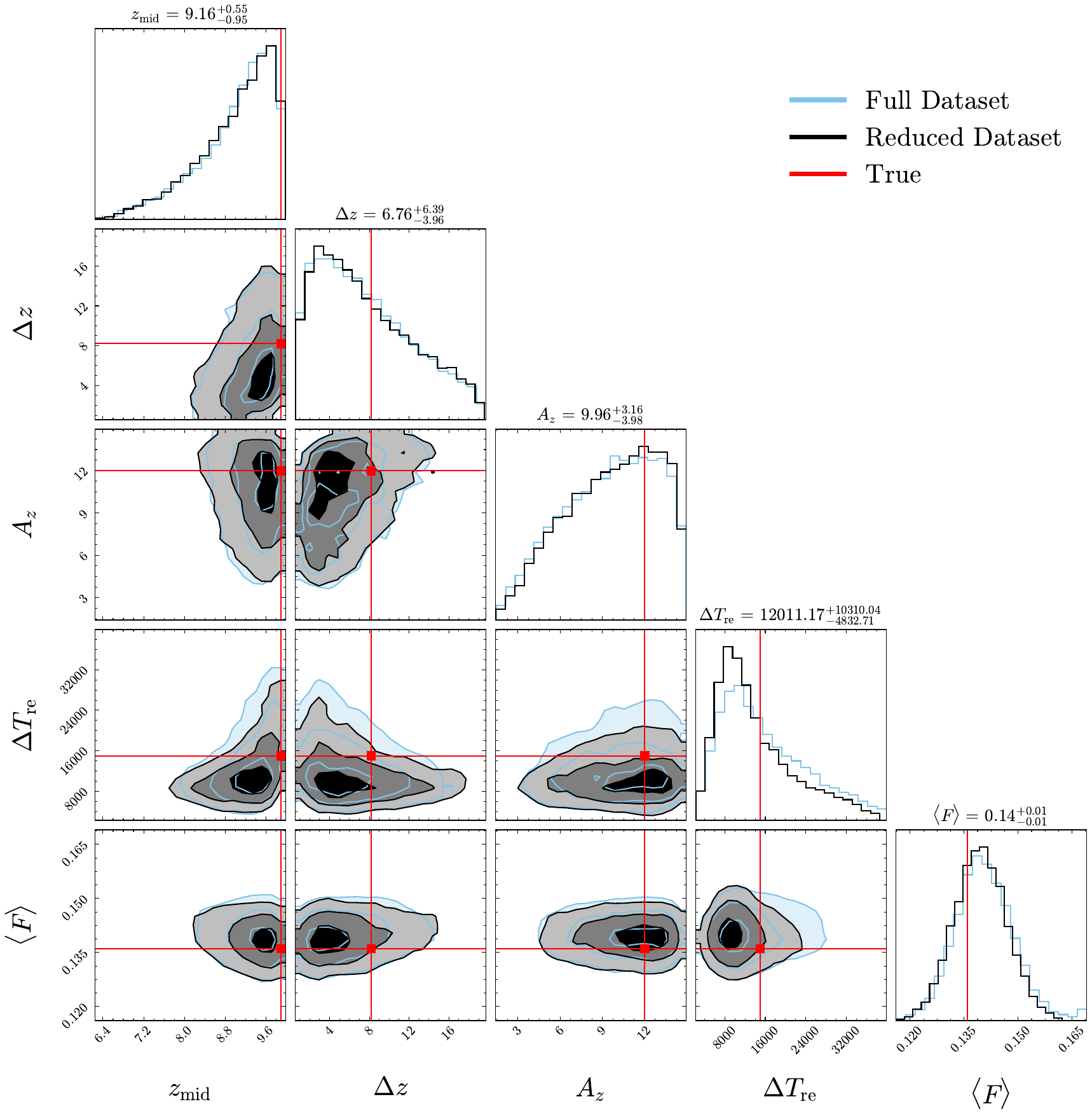}
    \caption{The posterior distribution obtained for the mock observations as the one shown in the right of Figure~\ref{fig:corner-plots}. The corner plot shows the posterior distributions obtained using the emulators trained on the reduced dataset (in black) overlaid on top of the posterior distributions obtained using the emulators trained with the complete dataset (in light blue). The true parameters of each mock observation are plotted in red. The median of the marginalized posteriors for each parameter is shown on top of the 1D histograms, along with the uncertainty of their estimation (given by the $16$-th and $84$-th quantiles). The true parameters for this model are $z_{mid}$ = 9.9, $\Delta z$ = 8.2, $A_{z}$ = 12.0, $\Delta T_{\text{re}}$ = 15000 K, $\langle F \rangle$ = 0.1360. As we can appreciate, the posterior obtained with the emulators trained on the reduced dataset match that of the emulators trained on the full dataset, with a slight overconfidence present in the estimates for $\Delta T_{\text{re}}$.}
    \label{fig:corner-plot-reduced}
\end{figure}

\begin{figure}
    \includegraphics[width=\columnwidth]{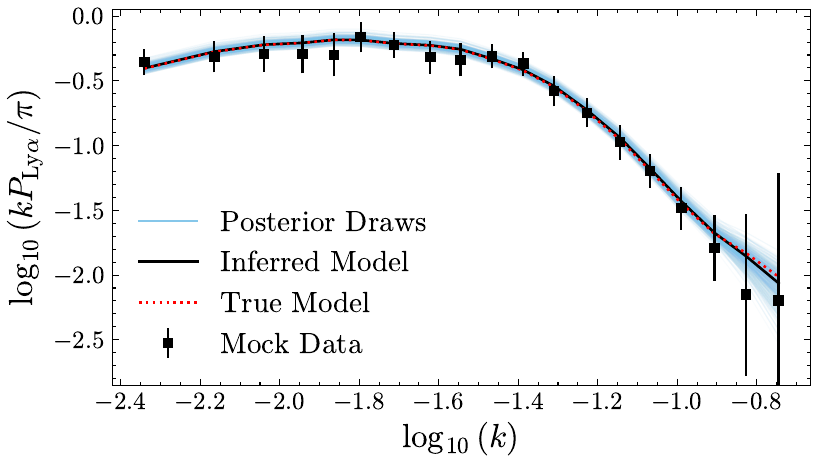}
    \caption{The inferred model plot for the posterior distribution in Figure \ref{fig:corner-plot-reduced}. For comparison, we show the true model (red dotted line), and 500 randomly selected posterior draws. We also plot the mock observations (square markers) along with their errors, which are taken to be equal to the square root of the diagonal elements of the covariance matrix corresponding to the true model respectively. }
    \label{fig:inferred-model-reduced}
\end{figure}

We then performed parameter estimation on the same set of 100 mock observations as in Section \ref{subsec:inference-test} to construct to check the statistical behavior of the posterior distributions obtained with these new emulators. Despite the significant increase in emulation error, visually inspecting the posterior distributions showed that the parameter estimation remained roughly the same. Figure~\ref{fig:corner-plot-reduced} shows an example of the posterior distribution obtained with the emulators trained with the reduced dataset, and compares it to the posterior obtained for the same mock observation using the emulators trained with the full dataset. Importantly, this is the same mock observation used for the corner plot shown on the right of Figure~\ref{fig:corner-plots}. Visually, there are no significant differences between these posterior distributions, aside for a slightly tighter estimate for $\Delta T_{\text{re}}$. Additionally, Figure~\ref{fig:inferred-model-reduced} shows the corresponding inferred model plot. Once again, visually inspecting this figure does not show any significant differences to the inferred model plot shown on the right of Figure~\ref{fig:inferred-models}. 

In addition, we performed the same inference test using 100 different posterior distributions obtained using the emulators trained on the reduced dataset. The resulting coverage plot showed that the statistical validity of the posterior distributions remained roughly the same, as seen in Figure \ref{fig:coverage-plot}, with a slight overconfidence seen at some confidence levels. 

\section{Discussion}

\subsection{Improving the reionization models}\label{subsec:improving-the-models}
As mentioned in Section \ref{subsec:simulation-dataset}, the set of simulations used in this work does not have enough resolution to resolve the small structure of the Ly$\alpha$ flux field, producing a suppression of power at small scales, as illustrated in Figure \ref{fig:power-spectrum-boera}. This prevents us from utilizing this dataset and our emulators to perform a scientifically valid inference. Instead, a set of simulations with a higher resolution of $2048^3$ voxels \citep[as used in][]{Doughty2025} is required, and is part of our future work.

Another limitation of our current modeling framework lies in the treatment of thermal feedback during reionization. In this work, we use $\Delta T_{\text{re}}$  to parametrize the maximum amount of heat injected into each gas cell upon its reionization, an approach found in the literature \citep[see e.g.][]{Doughty2023a, Doughty2025}. However, this prescription does not fully capture the physical processes that regulate the IGM’s thermal evolution. In reality, the temperature increase associated with reionization depends on the local properties of the ionization front, which in turn are sensitive to the overdensity of the gas and the interplay between photoheating and cooling processes \citep{Miralda1994, Davies2015}. 
More sophisticated treatments have shown that underdense regions experience stronger heating than overdense ones, related to more efficient Ly$\alpha$ cooling in dense gas and the speed of the ionization front \citep{Hirata2018, D'Aloisio2019}. Because the Ly$\alpha$ forest at $z > 5$ primarily probes underdense regions, such differences in thermal evolution can significantly alter small-scale structure by accelerating pressure smoothing in hotter gas. As a result, our simplified heat injection model may ultimately limit the accuracy of future constraints derived from statistics like the 1D flux power spectrum.

In addition, our current simulations do not include spatial fluctuations in the ultraviolet background (UVB). The absence of UVB fluctuations may have a particularly strong impact on certain Ly$\alpha$ forest statistics, such as the power spectra and the distribution of effective optical depths. Previous studies have shown that large-scale UVB fluctuations 
can boost clustering statistics across a range of physical scales, especially during the tail end of reionization \citep{Davies2015, Onorbe2019, Wolfson2023}. In future work, incorporating fluctuating UVB models will be an important step toward developing a more physically complete framework for reionization-era Ly$\alpha$ forest modeling.

\subsection{Emulator Performance}\label{subsec:emulator-performance-discussion}
After performing hyper-parameter tuning, both emulators were able to achieve a sub-percent
percent error on average across their respective test datasets. In particular, our power spectrum emulator shows a remarkable performance, consistent with other Ly$\alpha$ power spectrum emulators in the literature \citep{Cabayol-Garcia2023, Molaro2023, Bird2023}. Figure \ref{fig:power-spectrum-performance} shows that the relative error depends on $k$. We see an increase in the median error as well as its spread both at small and large values of $k$, corresponding to the large and small scale structure respectively. We argue that this is expected, since the different reionization fields caused by their respective histories cause different small and large scale distributions in the Ly$\alpha$ flux field. 

After performing hyper-parameter tuning, both the mean power spectrum and covariance matrix emulators were able to achieve a sub-percent and sub-five percent error on average across respectively. In particular, our power spectrum emulator shows a remarkable performance, consistent with other Ly$\alpha$ power spectrum emulators in the literature \citep{Cabayol-Garcia2023, Molaro2023, Bird2023}. Figure~\ref{fig:power-spectrum-performance} shows that the relative error depends on $k$. We see an increase in the median error as well as its spread both at small and large values of $k$, corresponding to the large and small scale structure, respectively. We argue that this is expected, since the different reionization fields caused by their respective histories induce distinct large and small scale fluctuations in the Ly$\alpha$ flux field. In particular, even at the largest emulation errors, the mean power spectrum emulator’s uncertainty remains below the expected observational uncertainties assumed in our foward modeling (see Section~\ref{subsec:forward-modeling}). This confirms that the emulator is well within the required precision to enable parameter inference.

We observe a similar behavior with our covariance matrix emulator. Figure \ref{fig:covar-performance} shows that most of the error budget comes from the elements in the covariance matrix that correspond to small scales. This is more clearly seen in Figure \ref{fig:covar-diag-performance}, where we can also appreciate that the spread in the relative error of the diagonal elements increases significantly for larger values of $k$. To our knowledge, this is the first time physically motivated, model-dependent covariance matrices have been emulated with neural networks for analysis using the Ly$\alpha$ forest. Consequently, we do not have other examples to directly compare the performance of our emulator. However, it is clear that it incurs in significantly larger errors compared to the power spectrum emulator. This is to be expected. Instead of predicting a relatively small array of values from our five parameters $\theta$ (each power spectrum has 19 values), the covariance matrix emulator is trying to predict a total of 190 different values (all elements of a single Cholesky factor) from the same 5-dimensional input, which is a significantly harder task. However, the covariance matrix emulator obtains a satisfactory performance that makes it usable for parameter estimation within our framework, allowing us to have a good approximation of varying model-dependent covariance matrices during sampling.

Reducing the training dataset produces a significant increase in emulation error, with the evaluation metrics for both emulators showing an increase of around an order of magnitude, especially at smaller scales. In the case of the mean power spectrum emulator, the emulation error becomes comparable to the expected uncertainty from noisy observations, which could be a potential concern. However, we argue that the emulator remains useful in this regime: although the errors approach the level of the observational noise, they generally remain below it across most $k$ modes. This suggests that the emulator does not dominate the total error budget at all scales, and can still be employed to perform parameter inference. On the other hand, the covariance matrix emulator's increase in error is equally concerning. Once again, this is caused by the difficulty of predicting a relatively large number of output values from a small number of inputs, a challenge that is also increased by the fact that different models differ more on smaller scales. 
However, despite the increase in emulation error, the performance seems to remain sufficient to perform parameter estimation, resulting in similar posterior distributions (as seen in Figure~\ref{fig:corner-plot-reduced}), with a similar statistical validity (as evidenced by the coverage plot shown in Figure \ref{fig:coverage-plot}), 
and gives us confidence that using a smaller dataset of high resolution datasets could be enough to train adequate emulators in future work. Moreover, we could make use of transfer learning techniques \citep[e.g.][]{Zhuang2019}, and make use of our low resolution dataset to first pre-train the new neural emulators, which could help alleviate the problem of having a small dataset of high resolution simulations.

\subsection{Inference Test}\label{subsec:inference-test-discussion}

A visual inspection of the posterior distributions obtained by using the trained emulators and applying our method shows that the 1D power spectrum of the Ly$\alpha$ forest can be used to provide some constraints of the reionization history at a $z\sim5$, as can be seen in Figure \ref{fig:corner-plots}. Additionally, the posteriors obtained with our method seem to be in agreement with the true models, as seen in Figure \ref{fig:inferred-models}. Performing inference on 100 randomly selected mock observations allowed us to test the statistical validity of our framework by constructing the coverage plot shown in Figure \ref{fig:coverage-plot}. As we can see, the calculated coverage is slightly under the ideal curve for most values of $\alpha$, with the ideal curve exceeding the Poisson error for all values higher than $\alpha\sim0.45$. This indicates that the posterior distributions obtained by our method present a slight overconfidence or bias. While not ideal, it seems to be in agreement with \citet{Wolfson2023, Wolfson2023mfp}. In those studies, the authors used the autocorrelation function of the Ly$\alpha$ forest at redshifts between $z=5.4-6.0$ and performed the same inference test by obtaining posterior distributions with a combination of nearest grid point (NGP) interpolation and MCMC sampling, and found an increase in overconfidence or bias with an increase in redshift, indicating that the assumption of a multivariate Gaussian likelihood function is not fully adequate, an issue that worsens at even higher redshifts. In \citet{Wolfson2023}, it is argued that this is because the Ly$\alpha$ forest is known to be a non-Gaussian random field. 
Thus, the assumption of a multivariate Gaussian likelihood is adopted expecting that averaging over all calculated powers at every $k$ mode will Gaussianize the resulting distribution of the values of the mean power spectrum for each model, as is expected from the central limit theorem. 

While a similar argument might explain our results, in principle, it is difficult to determine whether the observed overconfidence or bias can be fully attributed from the statistical assumptions in our framework or from the error introduced by our emulators. In the second case, it might be possible to propagate the emulator error and add the effects in the evaluation of our likelihood, as in \citet{Zhenyu2025}. 
\citet{Zhenyu2025, Grandon2022} used a method to determine to propagate the error that their Ly$\alpha$ autocorrelation function neural emulator produced, allowing them to create a term that is added to the covariance matrix during the evaluation of their likelihood. Following that approach, \citet{Zhenyu2025} obtained a better result in their inference test. Using a similar approach to propagate the error of our power spectrum emulator showed that there was virtually no improvement in our case, however. We conclude that this is because our power spectrum emulator performed remarkably well, introducing only small errors in its predictions. To our knowledge, no procedure has been developed to do a similar error propagation for the covariance matrix emulator, and we leave the development of such method to future work. Lastly, given that there is only small decrease in performance obtained with the emulators trained on the reduced dataset (which produced emulation errors of about an order of magnitude higher compared to the emulators trained on the full dataset), as the coverage plots in Figure \ref{fig:coverage-plot} show. Thus, we argue that most of the overconfidence seen in our posterior distributions come from the assumption of a Gaussian likelihood. A possible avenue to solve this issue, especially at higher $z$, might be found in the recent development of simulation-based inference (SBI) methods based on ML algorithms \citep{Cranmer2020}, which could allow us to construct approximate surrogates of the likelihood, letting go of the standard Gaussian assumption.

\section{Conclusions}\label{sec:conclusions}

In this work, we have presented a viable method that will eventually allow us to place new constraints on the history of the Epoch of Reionization (EoR) using measurements of the 1D power spectrum of the Ly$\alpha$ forest at $z \sim 5$. To achieve this, we constructed a dataset of forward-modeled simulations with user-defined reionization histories, allowing us to explore a broad range of reionization scenarios. We trained a neural network emulator for the Ly$\alpha$ power spectrum that performs with sub-2\% accuracy across all relevant scales, and, to our knowledge, developed the first neural emulator for model-dependent covariance matrices, achieving sub-1\% error in most cases. 
We combined these emulators with Hamiltonian Monte Carlo to accelerate inference. We demonstrated the success of this method by performing inference on a set of mock observations and validated the resulting posterior distributions using a statistical inference test, demonstrating that the posterior distributions are statistically well behaved, though slightly overconfident, likely caused by the approximate assumption of a multivariate Gaussian likelihood. Additionally, we showed that training with a drastically reduced dataset of a 100 simulations still yields emulators and posteriors with competitive performance. This suggests that the method is feasible even when applied to high-resolution simulations, where computational costs are prohibitive. We also developed a pipeline capable of automating some components of the method, including training, hyperparameter optimization, parameter estimation, and performance evaluation.

Given that the low resolution of our current simulations suppresses small-scale power in the Ly$\alpha$ forest, we are unable to apply our results to real observations. Furthermore, our reionization model relies on a simplified heat injection prescription and omits fluctuations in the UVB, both of which could affect key observables. Future work will focus on addressing these issues by incorporating a more physically motivated treatment of the radiative transfer components of the model, as well as inhomogeneous UVB fields to a set of high-resolution simulations. In addition, we plan to explore simulation-based inference methods that may better capture the non-Gaussian nature of the Ly$\alpha$ forest, especially at higher redshifts. Taken together, these improvements will allow this framework to provide new constraints on the EoR's history using the Ly$\alpha$ forest.

\section*{Acknowledgements}
We acknowledge insightful discussions with the ENIGMA group at UC Santa Barbara and Leiden University. JFH acknowledges support from the National Science Foundation under Grant No. 1816006. This research used resources of the National Energy Research Scientific Computing Center, which is supported by the Office of Science of the U.S. Department of Energy under Contract No. DE-AC02-05CH11231. This research further used resources of the Oak Ridge Leadership Computing Facility at the Oak Ridge National Laboratory, which is supported by the Office of Science of the U.S. Department of Energy under Contract No. DE-AC05-00OR22725.

\section*{Data Availability}
The code and simulation data presented in this paper are available for reproducing results upon reasonable request.



\bibliographystyle{mnras}
\bibliography{MAIN} 




\appendix

\section{Hyper-parameter Tuning}\label{app:hparam-tuning}
As explained in Section \ref{subsec:jax}, we applied the pipeline to perform hyper-parameter tuning for both emulators. For this study, we only varied the number of hidden layers $n_h$, the number of units for each layer $n_{u,i}$, the initial learning rate $l_{r}$, and the number of training epochs $n_e$. The best MSE losses were of $3.0\mathrm{e}{-6}$ and $1.2\mathrm{e}{-6}$ for the $\langle P_{\mathrm{Ly\alpha}} \rangle$ and $\Sigma_{P_{\mathrm{Ly\alpha}}}$ emulators respectively. Table \ref{tab:hparam-results} shows the hyperparameter values that produced the best results for both emulators.

\begin{table}
	\centering
	\caption{Results of the hyper-parameter tuning for both emulators. The combination of parameters listed in this table provided the best validation losses during training.}
	\label{tab:hparam-results}
	\begin{tabular}{ccccc} 
		\hline
		Emulator & $n_h$ & $n_{u,i}$ & $l_{r}$ & $n_e$ \\
		\hline
		$\langle P_{\mathrm{Ly\alpha}} \rangle$    & 5 & [8, 12, 16, 20, 22] & 0.0152 & 750 \\
            $\Sigma_{P_{\mathrm{Ly\alpha}}}$ & 5 & [25, 25, 25, 50, 50] & 0.0078 & 1250 \\
		\hline
	\end{tabular}
\end{table}

\section{Performance of Emulators trained on the reduced dataset}\label{app:reduced-performance}

\subsection{Emulator Performance}\label{app:reduced-performance-emulation}
\begin{figure}
    \includegraphics[width=\columnwidth]{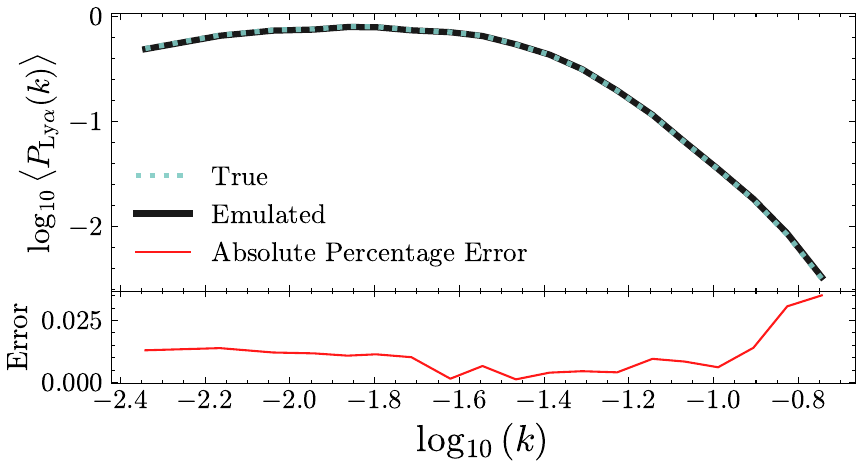}
    \caption{Example of an emulated power spectrum in the test dataset using the $\langle P_{\mathrm{Ly\alpha}} \rangle$ trained with the reduced training dataset, with an error in the 50th percentile. The emulated power spectrum is shown in the solid black line, while the true power spectrum is shown in the light blue dashed line. The bottom panel shows the absolute relative error as a function of $k$. }
    \label{fig:mean-emulation-example-reduced}
\end{figure}

As explained in Section~\ref{subsec:reduced-dataset}, we retrained both of our emulators using only 100 simulations for both the training and validation sets to test how their performance would be affected with a more realistic number of models. Unsurprisingly, when evaluated in the test dataset, these emulators both present an increase in error of about an order of magnitude, as seen in Figure~\ref{fig:power-spectrum-performance-reduced} and Figure~\ref{fig:covar-diag-performance-reduced}. Figure~\ref{fig:mean-emulation-example-reduced} shows an example of an emulated power spectrum in the test dataset, with an error in the 50th percentile. As we can observe, while the overall error has increased, this emulator is still performing exceptionally well, providing extra evidence that neural networks are well suited to predict the power spectrum as a function of our model parameters $\theta$. 

\begin{figure*}
    \centering
	\includegraphics[width=1.9\columnwidth]{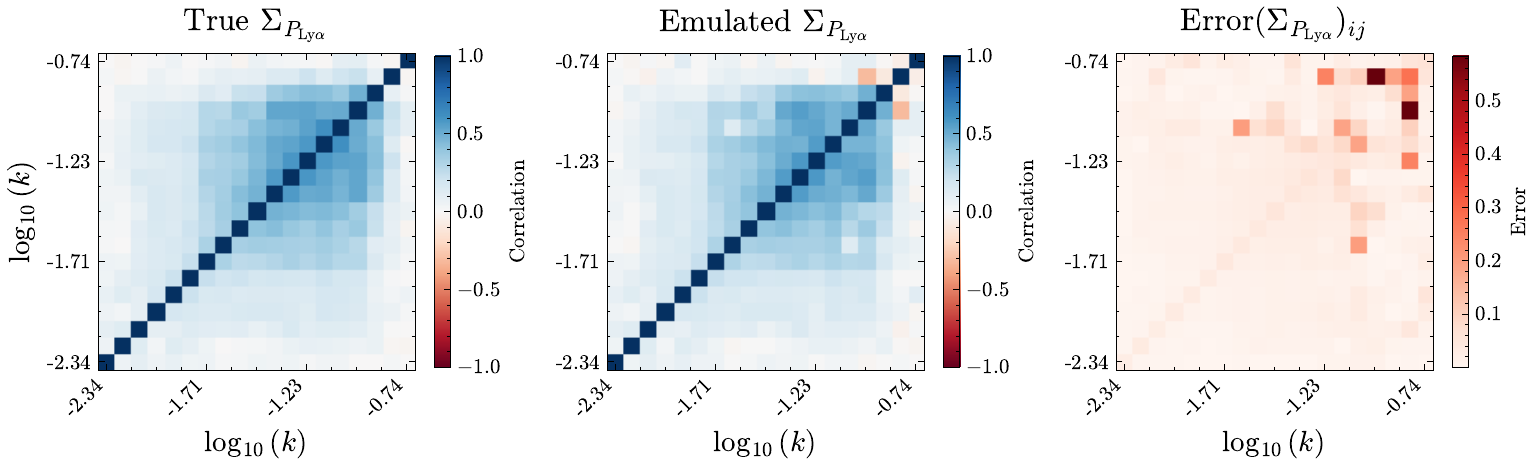}
    \caption{Example of an emulated covariance matrix obtained with the $\Sigma_{p_{\mathrm{Ly\alpha}}}$ trained with a reduced dataset, with an error in the 50th percentile. The true covariance matrix is shown in the left, while the emulated covariance matrix is shown in the middle. Both covariance matrices are shown as correlation matrices as described by Equation \ref{eq:correlation-matrix}. The heat map on the right shows the error as defined by Equation \ref{eq:covar_error} for this example.}
    \label{fig:covar-emulation-example-reduced}
\end{figure*}

On the other hand, Figure~\ref{fig:covar-emulation-example-reduced} shows an example of an emulated covariance matrix in the test dataset, with an error in the 50th percentile. The emulation error in this case has increased noticeably, with clear visual differences between the true and emulated covariance matrices. Additionally, it is evident that most of the error comes from diagonal elements that correspond to large $k$ modes. Once again, this is expected given the problem that predicting a large number of values from our five $\theta$ parameters is inherently a more difficult task. While not optimal, however, this emulator seems to have a performance that is sufficient enough to perform inference, as discussed in Section~\ref{subsec:emulator-performance-discussion}.

\bsp	
\label{lastpage}
\end{document}